\documentclass{svmult}
\usepackage{makeidx}  
\usepackage{graphicx}  
\usepackage{subfigure}   
\usepackage{mathptmx,helvet,courier}  
\usepackage{multicol}        
\usepackage[bottom]{footmisc} 

\def\rcgindex#1{\index{#1}}
\def\myidxeffect#1{{\bf\large #1}}

\makeindex             

\begin{document}
 \title*{Breather mobility and the PN potential  } 
 \subtitle{Brief review and recent progress} 
  \titlerunning{Breather mobility and the PN potential}
 \author{Magnus~Johansson \and  
  Peter~Jason
 } 
\institute{M~Johansson \at Department of Physics, Chemistry and Biology (IFM),
Link{\"o}ping University, SE-581 83  Link{\"o}ping, Sweden
  \email{mjn@ifm.liu.se}
  \and P~Jason
  \at
 Department of Physics, Chemistry and Biology (IFM),
Link{\"o}ping University, SE-581 83  Link{\"o}ping, Sweden
  } 
\authorrunning{M~Johansson and P~Jason}

\maketitle


\abstract{The question whether a nonlinear localized mode 
(discrete soliton/breather) can 
be mobile in a lattice has a standard interpretation in terms of the 
Peierls-Nabarro (PN) potential barrier. For the most commonly 
studied cases, the PN barrier for 
strongly localized solutions becomes large, rendering these essentially 
immobile. Several ways to improve the mobility by reducing the PN-barrier 
have been 
proposed during the last decade, and the first part gives
a brief review of  such scenarios in 1D and 2D. We then proceed to discuss 
two recently discovered novel mobility scenarios. The first example is 
the 2D Kagome lattice,  where the existence of a highly degenerate, 
flat linear band allows for  a very small PN-barrier and mobility of highly 
localized modes in 
a small-power regime. The second example is a 1D waveguide array in an active 
medium 
with intrinsic (saturable) gain and damping, where exponentially 
localized, travelling discrete dissipative solitons may exist as stable 
attractors. Finally, using the framework of an extended Bose-Hubbard model, 
we show that while quantum fluctuations destroy the mobility of slowly moving, 
strongly localized classical modes, coherent mobility of rapidly moving states
survives even in a strongly quantum regime. 
 }

\section{Introduction}
\label{johansson:sec:Intro}
 \rcgindex{\myidxeffect{A}!Annealing} 

The concept of a Peierls-Nabarro (PN) potential,  and a 
corresponding PN barrier, to describe the motion of a localized 
excitation in a periodic lattice has ancient roots. It originates 
in the work of Peierls from 1940~\cite{peierls1940}, 
later expanded and corrected by Nabarro~\cite{nabarro1947},
calculating the minimum stress necessary for moving a dislocation in a simple 
cubic lattice. A classical model for describing dislocation motion is the
Frenkel-Kontorova (FK), or discrete sine-Gordon, model~\cite{braun1998}, 
where dislocations appear as discretizations of the 
topological kink solitons of the continuum sine-Gordon 
equation. In the continuum limit, the system is Lorentz invariant so the
kink can be boosted to an arbitrary velocity without energy threshold. The 
lattice discreteness breaks the translational invariance and singles out 
two possible configurations for a stationary kink: a stable configuration 
centered in-between two lattice sites (``bond-centered'', ``inter-site'') 
and an unstable 
configuration centered at a lattice site (``site-centered'', ``on-site''). 
Defining the 
PN barrier as the minimum energy that must be added to a stable kink in order 
to translate it one lattice site, it becomes equal to the energy difference 
between the site-centered and bond-centered  kinks, since the kink must 
pass through a site-centered configuration in order to reach its next stable 
lattice position. If in addition one assumes that the kink travels very 
slowly and adiabatically through the lattice, one may employ a collective 
coordinate approach using the kink center as a collective coordinate. 
Calculating the kink energy as a function of its center then defines a PN 
potential as a continuous and periodic function of the lattice position, 
where stable positions appear as minima and unstable positions as maxima or 
saddles. See Ref.~\cite{braun1998} for more detailed discussions, and 
further references, concerning the PN potential for FK kinks. 

It is then highly tempting to carry over a similar reasoning for describing
the mobility also of nontopological lattice solitons, e.g., discrete breathers 
and discrete envelope and pulse solitons
(cf, e.g., Refs.~\cite{campbell1990,dauxois1993}). 
Indeed, as we will illustrate with 
many examples in the remainder of this chapter, such an approach is often 
very useful and has lead to much progress in understanding the conditions
for breather mobility in various models. 
However, some cautionary remarks may be in order 
before proceeding, in particlar for the reader more inclined towards
rigorous approaches. First, as was pointed out early 
by Flach and Willis~\cite{flach1994,flach1998}, a problem arises with the 
definition of a PN barrier/potential for discrete breathers in generic 
Hamiltonian lattices, since breathers come in continuous families and typically 
also have internal oscillation modes which may increase or decrease their 
energy. Thus, strictly speaking, there is no unique minimum energy needed for 
translating a breather one site since it depends on the internal breather
degrees of freedom, and therefore no well-defined PN barrier 
unless some additional constraint is imposed on the dynamics. This problem 
does not occur for kinks, since they carry topological charge and the 
stable kink is a global energy minimizer under the given boundary conditions. 
Second, the PN potential is defined assuming adiabatic (ideally infinitely 
slow) motion, and therefore the fact that a localized mode can be supplied 
with sufficient energy to overcome the PN barrier does not imply the
existence of \emph{exact} moving discrete solitons at \emph{finite} 
(possibly large) velocities. On the contrary: a localized mode travelling
through the periodic potential with a nonzero velocity will generate 
oscillations, which in the generic case will resonate with oscillation 
frequencies for linear waves. Thus, the motion causes radiation to be emitted, 
and the mode eventually slows down and/or decays. See
Ref.~\cite{flach2008a} for further discussion and references on this issue, 
and Ref.~\cite{pelinovsky2011} for a more mathematical approach. 

\begin{figure}[t] \begin{center}
\includegraphics[height=4.25cm, angle=0]{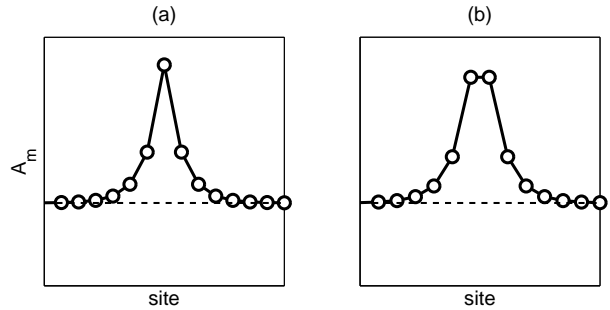}
\caption{ Illustrations of stationary DNLS breathers: (a) on-site; 
(b) inter-site. }
\label{fig:johansson-figure01}
\end{center}
\end{figure}

As mentioned, the first problem above may be overcome by imposing some 
additional constraint on the dynamics. As discussed by 
Cretegny and Aubry~\cite{cretegny1998,aubry2006}, 
a natural assumption would be that a breather 
moves at a constant total action, since for a time-periodic trajectory 
the action can be identified with the area inside a loop in phase space, 
which is conserved for any Hamiltonian system.
Thus, assuming adiabatic 
motion with a velocity much smaller than the oscillation frequency of the 
breather, the action should be at least approximately conserved also for a 
moving breather (see also Ref.~\cite{mackay2002} for a related approach, 
and Ref.~\cite{sepulchre2003} for a tutorial review).
In fact, for the very important class of Discrete 
Nonlinear Schr{\"o}dinger (DNLS) lattices~\cite{eilbeck2003a,kevrekidis2009}, 
which will be the main focus of this chapter, this 
statement is even rigorously true! The action then corresponds to the 
total norm (which, depending on the particular physical application of the 
model, may correspond e.g. to power, or particle number), which is a second 
conserved quantity of all DNLS-type lattices. In the particular case of 
a ``standard'' 1D DNLS chain with cubic, on-site nonlinearity 
(Eq.~(\ref{johansson:eq:xdnls}) below with $K_4=K_5=0$), it was realized
by Eilbeck already in 1986~\cite{eilbeck1986} 
(later rediscovered in Ref.~\cite{kivshar1993}) that
the proper definition of a PN barrier then
corresponds to comparing the energy of the on-site discrete soliton (which here 
is stable, Fig. \ref{fig:johansson-figure01} (a)) with the 
(unstable, Fig. \ref{fig:johansson-figure01} (b)) 
inter-site soliton \emph{at fixed norm}. 
He also concluded that for strong nonlinearity, corresponding to highly
localized solitons, the PN barrier grows proportionally to the nonlinearity 
strength, and therefore such solitons cannot be moved but are pinned to 
the lattice. In fact, it has later also been rigorously proven that stable DNLS 
solitons are global energy minimizers at fixed norm~\cite{weinstein1999}, 
which justifies the definition of the PN barrier as the minimum 
additional energy 
needed for translating the ground-state 
soliton one lattice site in slow, adiabatic motion. 
A very recent 
work~\cite{jenkinson2014} has 
also rigorously shown that for weak nonlinearity, when the DNLS soliton 
approaches the continuous NLS soliton, 
the PN-barrier becomes exponentially small in the discreteness parameter 
(lattice constant).

It should also be noted that although DNLS-type models have the non-generic
property of exact norm conservation, such models generically arise in
approximate, rotating-wave type, 
descriptions of the slow modulational, small-amplitude dynamics 
of more general nonlinear lattice models with anharmonic on-site 
(Klein-Gordon, KG) and/or intersite (Fermi-Pasta-Ulam, FPU) interactions. 
A separation of time-scales between fast, small-amplitude 
oscillations (e.g. breather frequency)
and slow modulations (e.g. breather movement) is a crucial ingredience in 
all such approaches, see, e.g., Refs~\cite{johansson2006,morgante2002} for 
discussion and further references. Thus, under these conditions, we should 
expect the Peierls-Nabarro potentials and barriers analyzed for DNLS models 
to also give good approximate descriptions of breather mobility in the 
corresponding KG/FPU lattices. 

After this very brief general review of the basic concepts of PN potential 
and barrier and their relation to breather mobility, the remainder of this 
chapter will focus on describing various ways to improve the mobility of 
strongly localized modes by reducing the PN-barrier, that have been proposed 
during the last decade. Sec.~\ref{johansson:sec:1D} discusses briefly 
the one-dimensional (1D) scenario, mainly within the framework of a 
DNLS model extended with inter-site nonlinearities. In 
Sec.~\ref{johansson:sec:2D}, we first give a general, short overview of 
different two-dimensional (2D) mobility scenarios that have been discussed 
in the literature, and then focus more particularly on the saturable DNLS model 
and the corresponding PN potential (Sec.~\ref{johansson:subsec:2Dsat}), 
and the Kagome lattice with mobile ``flat-band'' discrete solitons 
(Sec.~\ref{johansson:subsec:kagome}). Sec.~\ref{johansson:sec:gain} 
describes how an intrinsic gain may support exact localized 
travelling discrete dissipative 
solitons, and in Sec.~\ref{johansson:sec:QLC} we analyze the quantum 
mechanical counterparts to strongly localized moving modes, and discuss 
the conditions under which the classical PN potential concept has a meaningful 
quantum counterpart.


\section{PN-barriers and discrete soliton mobility in 1D}
\label{johansson:sec:1D}

As discussed above, for the ``standard'' DNLS equation, with a pure on-site, 
cubic nonlinearity, the energy difference between the stable, site-centered 
mode and the unstable, bond-centered mode is always nonzero and 
grows with increasing nonlinearity, 
and therefore strongly localized modes are highly immobile. Thus, in generic 
cases, we should expect PN potentials and barriers to be always nonvanishing. 
Exceptions occur for integrable models, such as the Ablowitz-Ladik 
discretization of the NLS equation, where the PN barrier strictly vanishes
since the model has continuous families of exact propagating soliton 
solutions~\cite{kivshar1993}. 

However, as 
was probably first noted for a cubic model with {\em inter-site} 
nonlinearities~\cite{oster2003}, also for some non-integrable models 
this energy difference may vanish in 
particular points when parameters are varied. The considered model was 
derived using a coupled-mode approach to describe stationary light propagation 
in an optical waveguide array embedded in a nonlinear Kerr 
material,
and after rescalings it takes the form of an extended DNLS equation,
\begin{eqnarray}
   \label{johansson:eq:xdnls}
    \I \dot{\Psi}_{n}= K_{2}(\Psi_{n-1}+\Psi_{n+1}) - \Psi_{n}|\Psi_{n}|^{2} 
\nonumber \\
    +2K_{4}\big(2\Psi_{n}(|\Psi_{n-1}|^{2}+|\Psi_{n+1}|^{2}) 
+ \Psi_{n}^{*}(\Psi_{n-1}^{2}+\Psi_{n+1}^{2})\big) \nonumber \\
    +2K_{5}\big(2|\Psi_{n}|^{2}(\Psi_{n-1}+\Psi_{n+1}) 
+ \Psi_{n}^{2}(\Psi_{n-1}^{*}+\Psi_{n+1}^{*}) +\Psi_{n-1}|\Psi_{n-1}|^{2} 
+ \Psi_{n+1}|\Psi_{n+1}|^{2}\big),
\end{eqnarray}
where the time-derivative in this context should be interpreted as a spatial 
derivative with respect to the longitudinal coordinate. 
For $K_4=K_5=0$, this is just the ordinary cubic DNLS model with 
nearest-neighbour coupling $K_2$ and on-site nonlinearity normalized to 1. The 
additional terms, whose strengths are detemined by parameters 
$K_4$ and $K_5$,  describe  two different types of 
nonlinear nearest-neighbour mode couplings, both resulting 
from the nonlinearity of the embedding medium. Like the ordinary DNLS 
equation, Eq.~(\ref{johansson:eq:xdnls}) has a standard Hamiltonian structure 
with conserved Hamiltonian (energy), 
\begin{eqnarray}
   \label{johansson:eq:xdnlsham}   H = \sum_n\Big[K_2\Psi_{n}\Psi_{n+1}^{*} 
- \frac{1}{4}|\Psi_{n}|^{4}
   { + K_{4}\left( 2|\Psi_{n}|^{2}|\Psi_{n+1}|^{2} + 
\Psi_{n}^{2}\Psi_{n+1}^{*\,2} \right)}\nonumber \\
    { + 2K_5\Psi_{n}\Psi_{n+1}\left(\Psi_{n}^{*\,2} 
+ \Psi_{n+1}^{*\,2}\right)}
      \Big] + \textrm{c.c.} \,
\end{eqnarray}
(c.c. denotes complex conjugate),
as well as conserved norm (power, excitation number),
 \begin{equation}
  P =  \sum_n |\Psi_n|^2 .
 \label{johansson:eq:norm} 
 \end{equation}
The fundamental discrete solitons (breathers) are, just as in the standard 
DNLS model, spatially localized stationary solutions with a purely harmonic 
time-dependence, 
\begin{equation}
\Psi_{n}(t)=u_{n}\E^{-\I\Lambda t},
 \label{johansson:eq:stat} 
\end{equation}
where the mode profiles $u_n$ generically can be chosen real and 
time-independent.

\begin{figure}[t] \begin{center}
\includegraphics[width=5.8cm, angle=0]{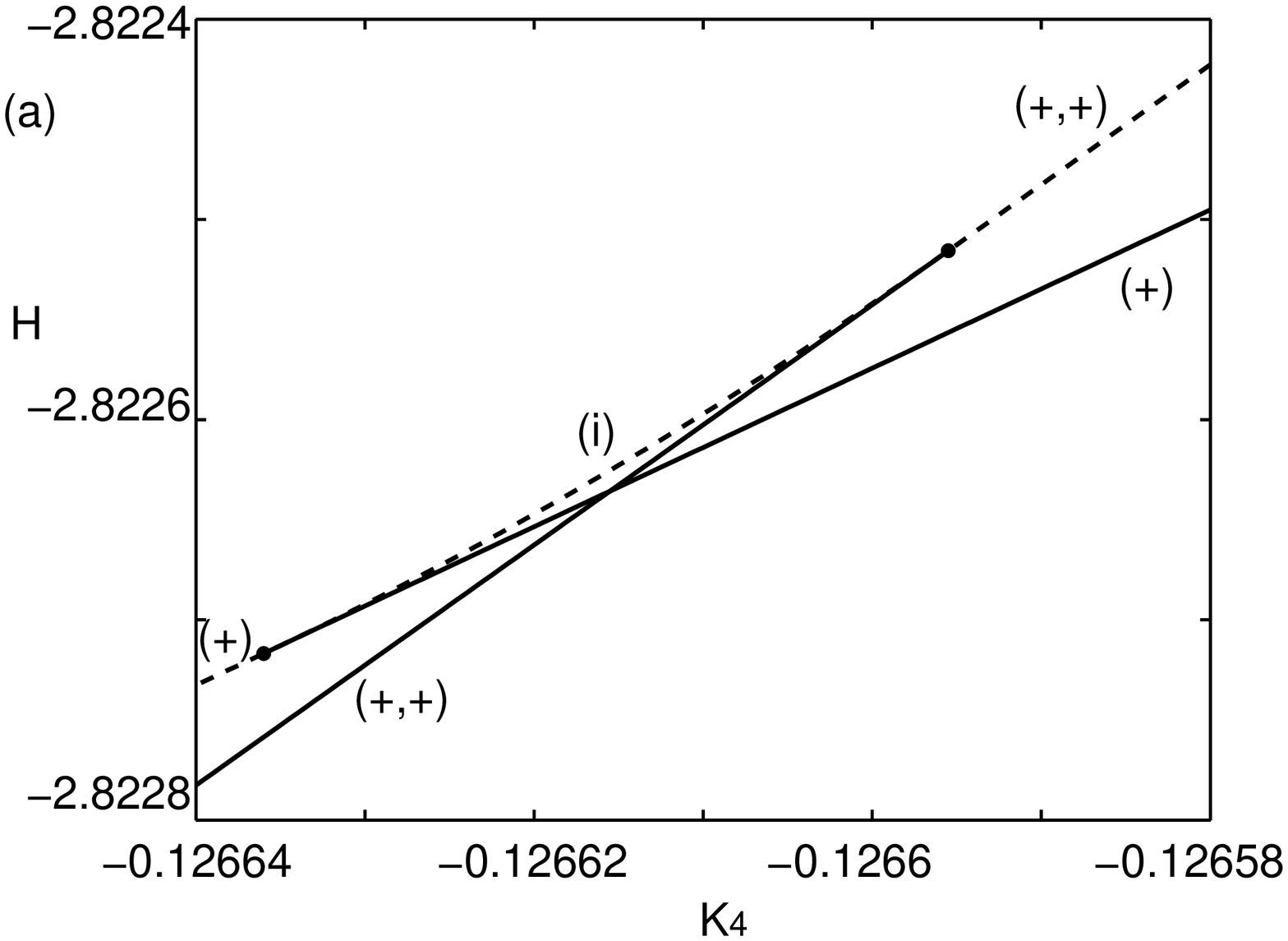}
\includegraphics[width=5.8cm, angle=0]{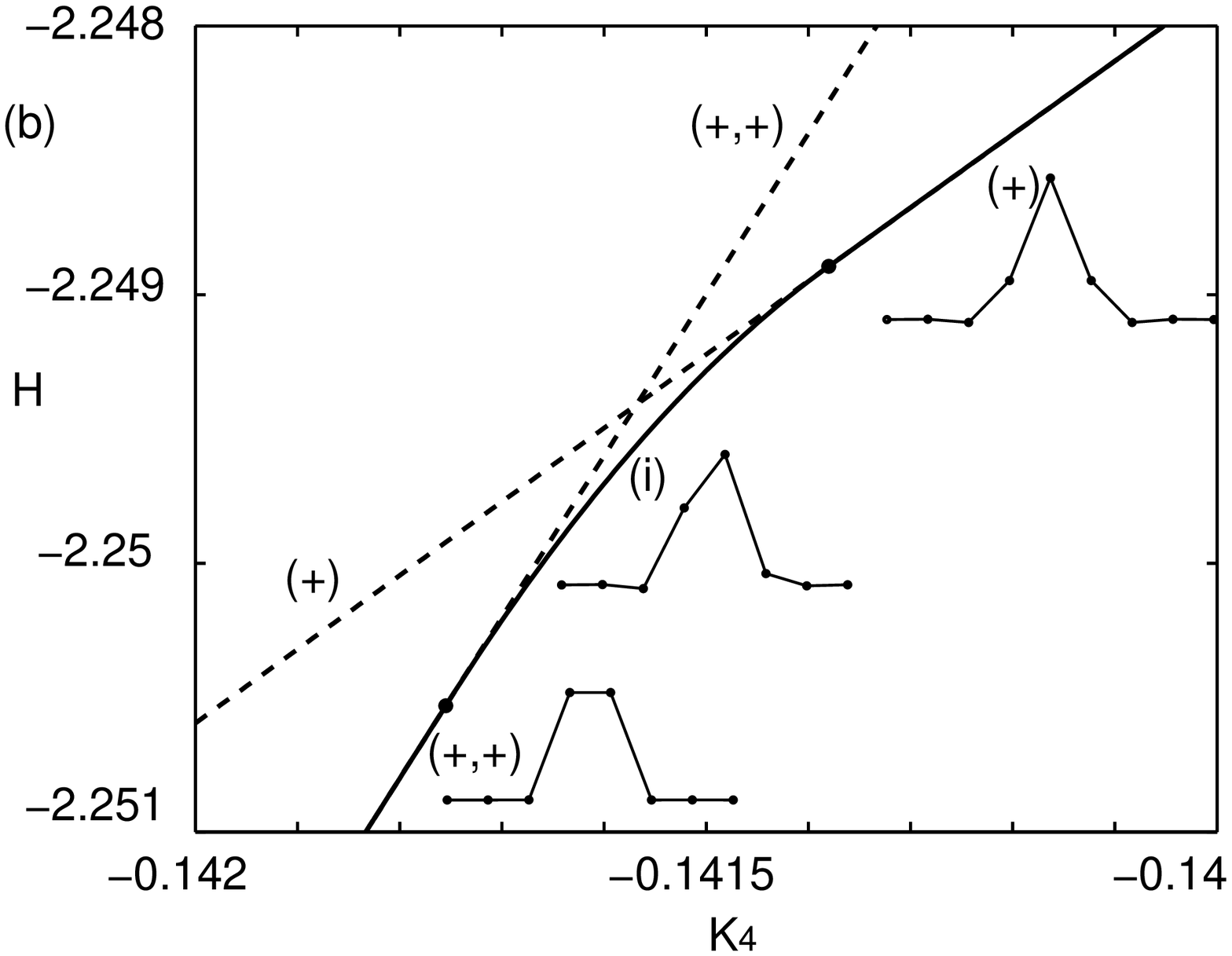}
\caption{Bifurcation diagrams for stationary discrete solitons in the
extended DNLS equation (\ref{johansson:eq:xdnls}) with $K_2=0.2$, 
having a constant norm (\ref{johansson:eq:norm}) $P=2$. The Hamiltonian 
(\ref{johansson:eq:xdnlsham}) is plotted as a function of the parameter 
$K_4$ for two different values of the other inter-site nonlinearity parameter,
(a) $K_5 = -0.18$, and (b) $K_5 = -0.1$. The symbols denote the three 
different types of solitons: on-site (+), symmetric inter-site (+,+), and 
asymmetric intermediate (i), with profiles (at $K_4=-0.1416$) indicated in (b).
Solid (dashed) lines denote linearly stable 
(unstable) solutions, and bifurcation 
points are indicated with dots. Figure adapted from Ref.~\cite{oster2003}. }
\label{fig:johansson-figure02}
\end{center}
\end{figure}

The vanishing, at specific values of $K_4$, of the energy difference between 
on-site and inter-site 
solutions having the same norm is illustrated in 
Fig. \ref{fig:johansson-figure02}, for two different values of $K_5$. 
As 
has been confirmed by studies of many other models (several of those 
to be described later in this chapter), this vanishing is generically 
associated with a \emph{stability exchange} between the on-site and inter-site 
modes, appearing through bifurcations with a family of \emph{intermediate}, 
asymmetric 
stationary solutions, connecting the two types of symmetric solutions and 
``carrying'' the (in)stability between them. In fact, such a scenario 
for enhanced mobility had been originally described  by Cretegny and 
Aubry~\cite{cretegny1998,aubry2006} for breathers 
in a KG model with a Morse potential. 

One very important point to note here is, that close to such points, the 
\emph{true} 
PN barrier (defined, as discussed in Sec.~\ref{johansson:sec:Intro}, 
as the minimum energy needed for a lattice translation of a stable soliton) is 
\emph{not} equal to the energy difference between the on-site and inter-site 
modes, but generally \emph{larger} since energy is needed to pass also 
through the intermediate stationary solution. An analogous scenario 
has been known for a long time to appear for kinks in a modified FK model 
with a deformable substrate potential~\cite{peyrard1982}. 

\begin{figure}[t] \begin{center}
\includegraphics[width=12 cm, angle=0]{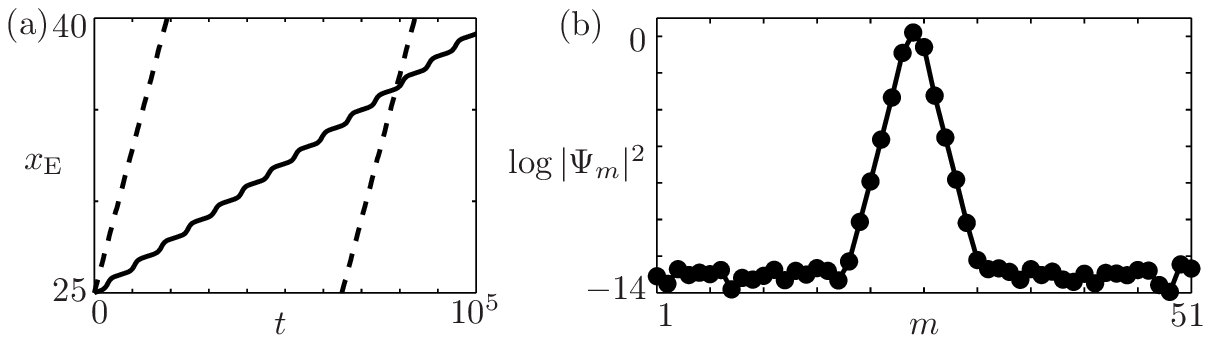}
\caption{Propagating excitation very close to a transparent point in the
extended DNLS equation (\ref{johansson:eq:xdnls}) with $K_2=0.2$, 
$K_4=-0.1316$, $K_5=-0.1470$, $P=2.01464$, and $H=-2.61566$. 
(a) shows the motion of the center of energy in a 51-site lattice with 
periodic boundary conditions. The (unstable) on-site stationary solution 
is perturbed by a phase gradient $\E^{\I k m}$ with $k=10^{-4}$ (solid) and 
$k=10^{-3}$ (dashed) ($m$ here denotes site index). 
The non-constancy of the velocity for the very 
slowly moving solution (solid) is a result of the finite numerical precision
when determining the transparent point; the remaining PN barrier is 
of the order of $10^{-7}$. (b) shows a snap-shot of the slowest 
excitation at $t=1000$.
Note the very small but non-vanishing tail, which is larger than the numerical 
accuracy and thus a result of the emitted radiation during the motion. 
From Ref.~\cite{oster2007}. }
\label{fig:johansson-figure03}
\end{center}
\end{figure}

Another important 
point is to note the qualitative difference between the two scenarios 
in Fig. \ref{fig:johansson-figure02} (a) and (b): in (a), the intermediate 
solution is unstable (energy max) and the on-site and inter-site solutions are 
simultaneously stable in the stability exchange region, while in (b) both 
symmetric solutions are unstable and the intermediate solution is stable 
(energy min). Thus, by varying also the second parameter (here $K_5$), it is 
possible to tune the existence regime for the intermediate solution, and even 
to make it \emph{vanish} at certain points! At such points, termed 
``transparent points'' in Ref.~\cite{melvin2006} (for a different model 
with saturable on-site potential to be discussed below), the PN barrier 
is indeed truly zero and a single family of translationally invariant 
stationary states 
having the same energy and norm must exist, with a free parameter 
corresponding to the position of
the center of energy. As 
elaborated for the model in Ref.~\cite{melvin2006} 
(see also Ref.~\cite{champneys2011} for further discussion and references), 
travelling waves do indeed 
bifurcate from stationary solutions at such exceptional points, but 
radiationless mobility is possible only at ``special'', nonzero, velocities. 
For generic small velocities, resonances with linear oscillations 
causing radiation cannot be avoided 
and so the mobility may be extremely good, but not perfect. An illustration 
of the almost perfect mobility for the model (\ref{johansson:eq:xdnls}) 
very close to a transparent point was given by 
{\"O}ster~\cite{oster2007} and is reproduced in 
Fig. \ref{fig:johansson-figure03}.
To our knowledge, it has not yet been investigated 
whether exceptional velocities with radiationless mobility exist also 
for Eq.~(\ref{johansson:eq:xdnls}). It was also noted in Ref.~\cite{oster2007} 
that even though the Hamiltonian and norm are independent of the location 
of the center of energy for the family of stationary solutions at the
transparent point of Eq.~(\ref{johansson:eq:xdnls}), the oscillation 
frequency $\Lambda$ is not. It was also found earlier, for slowly moving 
breathers in the stability exchange regime
of the KG chain with Morse potential, that the local oscillation frequency is 
not constant but varies with the location in the unit cell 
\cite{cretegny1998,aubry2006}.

As mentioned above, a similar scenario appears also for a DNLS model 
with a \emph{saturable} nonlinearity, which can be obtained from 
Eq.~(\ref{johansson:eq:xdnls}) by removing the intersite nonlinear terms 
($K_4 = K_5 = 0$) and replacing the cubic on-site nonlinearity with the term 
$\beta \Psi_n / (1 + | \Psi_n|^2)$. The Hamiltonian can then (after 
a trivial gauge transformation) be written as 
$H= \sum_n [\beta \ln (1+|\Psi_n|^2) + K_2 |\Psi_{n-1}-\Psi_n|^2]$. This
model is often used for describing spatial solitons in photorefractive 
waveguide arrays, and as was originally
discussed by Had{\v z}ievski et al.\ in 2004~\cite{hadzievski2004}, there are 
multiple points where the energy difference between on-site and inter-site 
discrete solitons vanish, and a very good mobility was observed. Many works 
have followed discussing various properties of these modes, of which we here 
just mention a few. Khare et al~\cite{khare2005} obtained analytical 
solutions for a complete family of intermediate solutions, Cuevas and 
Eilbeck~\cite{cuevas2006} studied discrete soliton interactions, 
Melvin et al.~\cite{melvin2006} found, as mentioned above, radiationless 
travelling waves at ``special'' velocities, and Naether 
et al.~\cite{naether2011b} analyzed the PN potential landscape in the 
stability exchange regimes using a constraint method to be described in the 
next section. We will also return to discuss the 2D version of the 
saturable DNLS and its mobility properties in the next section.




\section{Discrete soliton (breather) mobility in 2D}
\label{johansson:sec:2D} 
As is commonly known, mobility in 2D is ``normally'' much worse than in 1D, 
at least when the effective nonlinearity is cubic as in the standard 
DNLS-type models. 
As discussed e.g.\ in Ref.~\cite{christiansen1996}, the reason for this can be 
traced to the fact that in the continuum limit, 2D NLS solitons are
unstable and may undergo collapse into a singularity spike in a finite time. 
In a lattice, the 
strict mathematical collapse is impossible due to norm conservation, but 
instead a ``quasicollapse'' scenario appears where broad discrete 
(stationary or moving) solitons are transformed into highly localized and
strongly pinned modes~\cite{christiansen1996}. Moreover, it is important 
to note that, in contrast to the 1D case with cubic nonlinearity where 
the norm goes to zero in the small-amplitude (continuum) limit, 
the norm of 2D small-amplitude discrete solitons goes to a finite, 
nonzero value. The
consequence is the existence of an \emph{excitation threshold}, i.e., a 
minimum value of the norm below which no localized excitation exists, 
which has been rigorously established in Ref.~\cite{weinstein1999} (see 
also the recent discussion in Ref.~\cite{jenkinson2014}).

However, some notable exceptions to the general folklore ``mobility is
bad in 2D lattices'' has been known for some time, and we here try to briefly 
exemplify different physical situations where good 2D mobility of 
localized modes has been observed, and explain why the scenarios differ 
from the generic one described above.

(i) Moving breathers in vibrational lattices with several degrees of freedom, 
(e.g.\ longitudinal and transversal), such as the two-component hexagonal 
lattice used by Mar{\'i}n, Eilbeck and Russell~\cite{marin1998} to simulate 
the motion of quasi-one-dimensional ``quodons'' along certain directions 
in a mica-like structure. In this case, the vibrational direction singles 
out a preferred direction for the breather which breaks the 2D lattice 
symmetry. As a consequence, along ``suitable'' lattice directions 
the breather may strongly deform and become elongated along one 
direction and compressed along the other. Thus it should behave essentially 
as a 1D small-amplitude breather in this direction.

(ii) Moving 2D polarons 
have been observed in electron-phonon coupled lattices with anharmonic 
vibrational degrees of 
freedom, such as the Holstein model with saturable anharmonicity in 
Ref.~\cite{zolotaryuk1998}. In this case, the effective nonlinearity 
in the semiclassical dynamics is 
no longer purely cubic but saturable, and as will be disussed in detail 
below (Sec.~\ref{johansson:subsec:2Dsat}), 
such nonlinearity allows for stable, mobile localized modes also in 2D. 
Recently, another example of a system which may support mobile 2D polarons was 
given in Ref.~\cite{mozafari2013}, where a molecular lattice having 
both intra- and inter-molecular harmonic degrees of freedom was considered, 
and the electron-lattice coupling was assumed to have as well an on-site 
(Holstein) as an inter-site (Peierls) part. By tuning the relation between 
these two couplings suitably, mobile polarons were observed in a rather 
narrow parameter window. Thus, this mechanism of enhanced mobility by 
competing effective on-site and inter-site nonlinearities is reminiscent 
of the scenario for the 1D extended DNLS model (\ref{johansson:eq:xdnls}). 

(iii) Strongly anisotropic lattices, with essentially 1D mobility in the 
strong-coupling direction only. Typically these states are elongated,
and strongly localized in the weak-coupling direction only, where they are 
not mobile. For anisotropic DNLS models, this scenario was described in 
detail in Ref.~\cite{gomez-gardenes2006}. A related example is the 
``reduced-symmetry'' gap solitons~\cite{fischer2006}, where, 
although the lattice itself is isotropic, 
there is an effective anisotropy induced by anisotropic dispersion
at a band edge of a higher band (e.g.\ $p$-band). This 
scenario was analyzed in detail within a discrete coupled-mode approach
in Ref.~\cite{johansson2009}, where each lattice 
site is assumed to support two orthogonal, degenerate modes of dipolar 
character. With this approach, the mechanism of symmetry breaking thus 
becomes analogous to that of the two-component vibrational lattice discussed 
in (i) above, with the orientation of the local dipole corresponding to 
the direction of local lattice vibration in (i).

(iv) Moving, stable, small-amplitude 
``quasi-continuous'' breathers (wide relative to the lattice spacing) 
were found in (scalar) 
2D FPU-type lattices, square~\cite{butt2006} as well as 
hexagonal~\cite{butt2007}. Although a standard 
continuum approximation to lowest order yields a cubic NLS equation, 
where stable localized solutions do not exist as discussed above, 
their existence in the 2D FPU-lattice was explained by incorporating 
higher-order dispersive and nonlinear terms as perturbations, which under 
certain conditions could lead to stabilization. A similar effect was 
seen for moving solutions of very small amplitude in the cubic on-site DNLS 
equation~\cite{arevalo2009}. Essentially, the velocity makes the effective 
dispersion of the corresponding continuum NLS model anisotropic, resulting 
in a deformation of broad solitons which may move for rather long distances 
without collapsing or trapping. However, it was noted in 
Ref.~\cite{arevalo2009} 
that also these moving quasi-continuous solutions are weakly unstable and 
slowly decaying through dispersion in the DNLS lattice. 

(v) Systems with non-cubic effective nonlinearities in the 
equations of motion. For a quadratic 
nonlinearity, there are no collapse instabilites in the continuum limit 
in 2D, and no excitation threshold for discrete solitons. 
Thus, as shown in Ref.~\cite{susanto2007} for a 2D lattice with 
second-harmonic generating nonlinearity, 
the PN barrier for small-amplitude, weakly localized solutions 
may be small enough for good mobility in arbitrary lattice 
directions. The case with saturable 
nonlinearity~\cite{vicencio2006,naether2011a} has already been mentioned 
above in (ii) and will be disussed in  detail 
in Sec.~\ref{johansson:subsec:2Dsat} below. A similar mobility 
scenario 
appears also for a 2D DNLS model with competing (i.e., of 
different sign) cubic and quintic on-site nonlinearities~\cite{chong2009} 
(resulting 
e.g.\ from taking into account only the lowest-order terms in a Taylor 
expansion of a full saturable potential). 

(vi) Systems with flat, i.e. dispersionless, linear bands, such as the 
DNLS model for a 
Kagome lattice~\cite{vicencio2013} to be described in more details in 
Sec.~\ref{johansson:subsec:kagome}. In 
this case, the absence of linear dispersion implies that discrete solitons 
bifurcating from the flat band cannot be described by a continuous NLS 
equation, and therefore they are not prone to collapse instabilities. Instead, 
they bifurcate from localized linear modes with zero norm threshold 
also for cubic nonlinearities, and 
small-amplitude solutions can 
be movable while being still strongly localized due to the smallness of 
the PN-barrier in some regime.


\subsection{Discrete soliton mobility in the 2D saturable DNLS model}
\label{johansson:subsec:2Dsat}

The mobility properties of discrete solitons in the 2D DNLS model 
with a saturable on-site nonlinearity were first described in 
Ref.~\cite{vicencio2006} and further analyzed in Ref.~\cite{naether2011a}. 
With the notation from Ref.~\cite{vicencio2006}, describing spatial 
solitons in a photorefractive waveguide array, the dynamical equation 
takes the form
\begin{equation}
i \frac{\partial u_{n,m}} {\partial \xi} + \Delta u_{n,m} -\gamma\frac{u_{n,m}}
{1+|u_{n,m}|^2} = 0 ,
\label{johansson:eq:sat} 
\end{equation}
where $\xi$ is the normalized propagation distance along the waveguides 
(playing the role of the time coordinate in the standard Hamiltonian 
framework), $u_{n,m}$ describes the 
(complex) electric-field amplitude at site $\{n,m\}$, and 
$\Delta$ represents the 2D 
discrete Laplacian, 
$\Delta u_{n,m} \equiv u_{n+1,m}+u_{n-1,m}+u_{n,m+1}+u_{n,m-1}$, defining 
the linear interaction between nearest-neighbour waveguides. The two 
conserved quantities for Eq.~(\ref{johansson:eq:sat}) are the Hamiltonian 
(energy)
%
\begin{equation}
H =-\sum_{n,m} \left[ (u_{n+1,m}+u_{n,m+1})\ u_{n,m}^{*}
-\frac{\gamma}{2} \ln (1+|u_{n,m}|^2) +\textrm{c.c.} \right], 
\label{johansson:eq:H}
\end{equation}
and the power (norm)
\begin{equation}
P=\sum_{n,m} |u_{n,m}|^2. 
\label{johansson:eq:P}
\end{equation}

\begin{figure}[t] \begin{center}
\includegraphics[width=8 cm, angle=0]{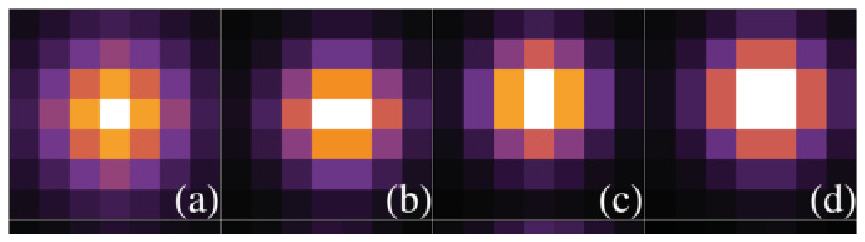}
\includegraphics[height=8 cm, angle=270]{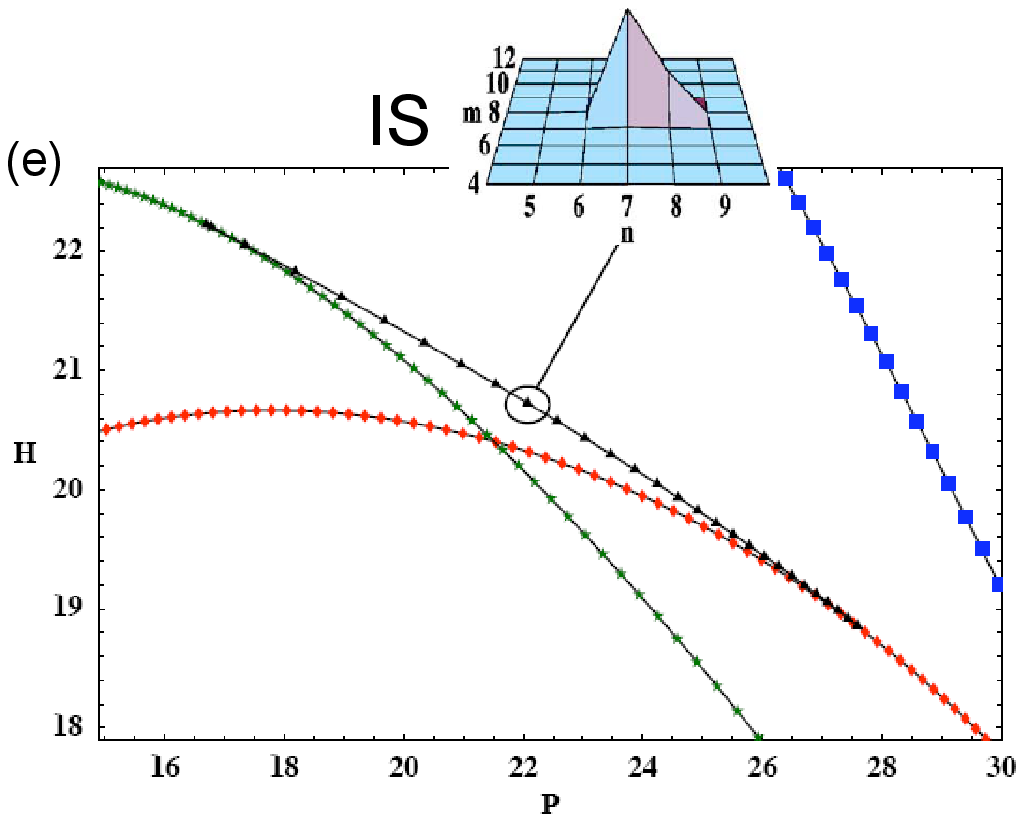}
\caption{Upper figures: examples of spatial profiles for the fundamental 
symmetric stationary solutions of the 2D saturable DNLS model 
(\ref{johansson:eq:sat}): (a) 1-site, (b) 2-site horizontal, (c) 
2-site vertical, and (d) 4-site. 
From Ref.~\cite{naether2011a}.
Lower figure (e): Bifurcation diagram showing the exchange of 
stability when $\gamma=10$ 
for increasing power from stable 1-site (red diamonds) to stable 
2-site (green stars) via an unstable intermediate solution (IS, black 
triangles) with profile indicated as inset. The 4-site solution 
(blue squares) is unstable in this regime. Adapted from 
Ref.~\cite{vicencio2006}. }
\label{fig:johansson-figure04}
\end{center}
\end{figure}

As illustrated in Fig.~\ref{fig:johansson-figure04}, 
there are three different types of fundamental symmetric stationary 
solutions, $u_{n,m}(\xi)=U_{n,m} \E^{\I \lambda \xi}$, 
which  will here be termed 1-site, 2-site, and 4-site modes, 
respectively, referring to the number of sites sharing the main peak of 
the modes. (These modes go under various other names in the literature, 
e.g., in Ref.~\cite{jenkinson2014} they are termed vertex-, bond- and 
cell-centered bound states, respectively.)
Note that there are two, degenerate, 2-site modes, horizontal 
and vertical. 
In bifurcation scenarios similar to that discussed for the 1D cases above, 
1-site, 2-site and 4-site modes may exchange their stability properties under 
variation 
of the parameters $\gamma$ and $P$, and as described in \cite{vicencio2006}, 
mobility then appears along axial directions in certain parameter regimes.
An example of the stability exchange between a 1-site and a 2-site mode, 
with the appearance of an unstable asymmetric intermediate solution (IS), 
is illustrated in the lower part of  Fig.~\ref{fig:johansson-figure04}.

In order to better understand the conditions for mobility in the various 
regimes, a numerical method was implemented in Ref.~\cite{naether2011a} 
for calculating the full PN potential
landscapes, showing the variation of the energy with the center of mass 
for localized solutions. The basic idea builds on the standard 
Newton-Raphson (NR) scheme for calculating stationary soliton solutions 
to the equations of motion (\ref{johansson:eq:sat})  (see, e.g., 
Ref.~\cite{flach2008a}), but imposes two additional constraints in order 
to fix the center of mass of the soliton horizontally and vertically:
\begin{equation}
X\equiv\frac{\sum_{nm} n |u_{n,m}|^2}{P}\ \ \ \ \ \ 
\textrm{and}
\ \ \ \ \ \ Y\equiv\frac{\sum_{nm} m |u_{n,m}|^2}{P}\ .
\label{johansson:eq:xy}
\end{equation}
Technically, this is implemented by eliminating the equations for two 
specific sites, chosen close to, but away from, the soliton center site, 
from the NR iteration, and instead determining the amplitudes for these 
sites (which can be chosen real and positive for the fundamental solitons)
from the constraint conditions (\ref{johansson:eq:xy}). (See 
Ref.~\cite{naether2011a} for further details and discussions about optimal 
choices of constraint sites.) Starting then from a stationary solution, 
e.g., a 1-site solution with center of mass at some lattice site
$(X,Y) = (n_c, m_c)$, we may proceed with a numerical continuation 
(at fixed power $P$) by increasing
adiabatically e.g. $X$ in the constraint (\ref{johansson:eq:xy}), until 
we end up at the horizontal 2-site solution centered at 
$(X,Y) = (n_c+1/2, m_c)$. From there, we may continue by increasing 
$Y$ adiabatically towards the 4-site solution at $(X,Y) = (n_c+1/2, m_c+1/2)$. 
Assuming that all NR iterations converge, it should be clear that the 
continuation could be done in any direction, and that we can also continue, 
e.g., with increasing $Y$ for any $X$ between $n_c$ and $n_c+1/2$. 

By 
calculating the energy (\ref{johansson:eq:H}) for each converged, 
constrained solution obtained from a sweep over the full area 
$n_c \leq X \leq n_c+1/2,\  m_c \leq Y \leq m_c+1/2$, we obtain a smooth 
PN potential surface if the continuation is smooth everywhere. A good 
mobility should then be expected if there are directions where these 
surfaces are smooth and flat. Note that only the {\em local extrema} 
of these surfaces may correspond to true stationary solutions of the 
{\em unconstrained}  system  (\ref{johansson:eq:sat}): stable solutions 
correspond to minima and unstable solutions to maxima 
or saddles.\footnote{A cautionary remark may be in order: if the constraint 
sites are 
not properly chosen, 
the method may reach different stationary solutions, or no stationary 
solutions at all~\cite{naether2011a}, and therefore yield 
different energy landscapes not related to the PN potential between the 
fundamental 1-site, 2-site and 4-site modes.} This type of method for 
calculating PN potentials was originally proposed for 1D breathers by 
Cretegny and Aubry~\cite{cretegny1998,aubry2006}, and similar methods 
were implemented e.g. for 1D kinks in KG chains~\cite{savin2000}, and 
applied to surface modes in the 1D DNLS model~\cite{molina2006}. 

An extensive discussion about the nature of the obtained PN surfaces in 
different parameter regimes, and the associated mobility properties, 
was given in Ref.~\cite{naether2011a}; here we just give a brief summary 
and show sample results for two particularly interesting regimes when 
$\gamma=4$, where smooth, complete surfaces were found for all values 
of the power $P$.  
In the low-power regime, the surfaces have single minima corresponding to 
the stable 1-site modes, saddle points corresponding to the unstable 2-site 
modes, and maxima corresponding to the likewise unstable 4-site modes. This 
ordering of energies for the stationary solutions is the same as for the 
ordinary (cubic) DNLS model (see, e.g., Ref.~\cite{jenkinson2014}), which 
could be expected since a small-power expansion of the saturable nonlinearity 
yields a cubic term to lowest order. However, as discussed above, 
for the cubic 2D DNLS model, stable stationary solutions are not mobile 
due to the large power excitation thresholds and narrowness of the 
stable solutions.  The effect of the saturability is to lower the excitation 
thresholds for all three stationary solutions~\cite{vicencio2006}, allowing for 
the existence of a regime of relatively low power with broader stable modes 
having
improved mobility~\cite{vicencio2006,naether2011a}. In terms of PN potentials,  
this results in smooth, complete 
2D surfaces generated from the constrained NR method, which 
could not be obtained for the cubic DNLS model~\cite{naether2011a}.

\begin{figure}[t] \begin{center}
\includegraphics[width=5.5 cm, angle=270]{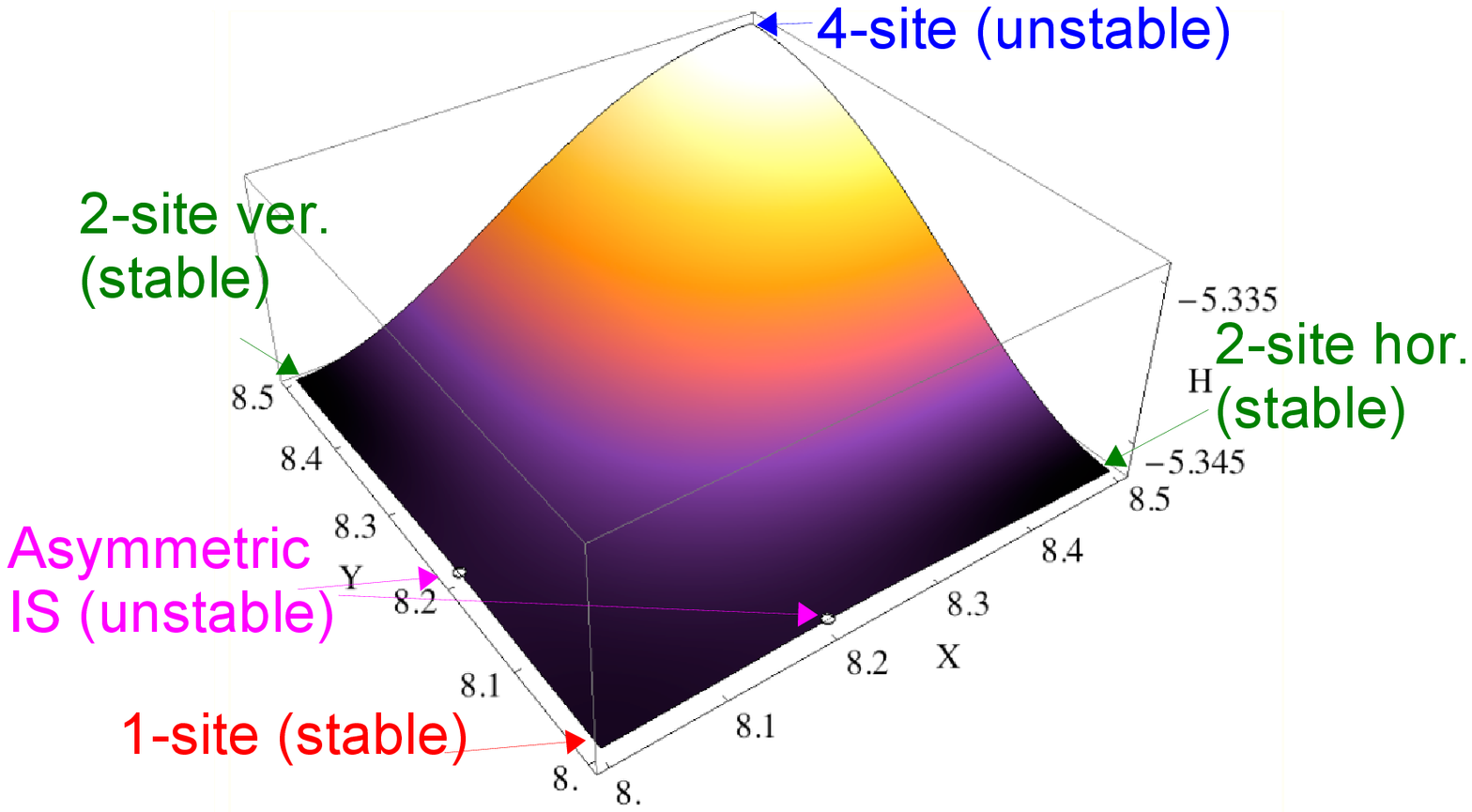}
\includegraphics[width=5.5 cm, angle=270]{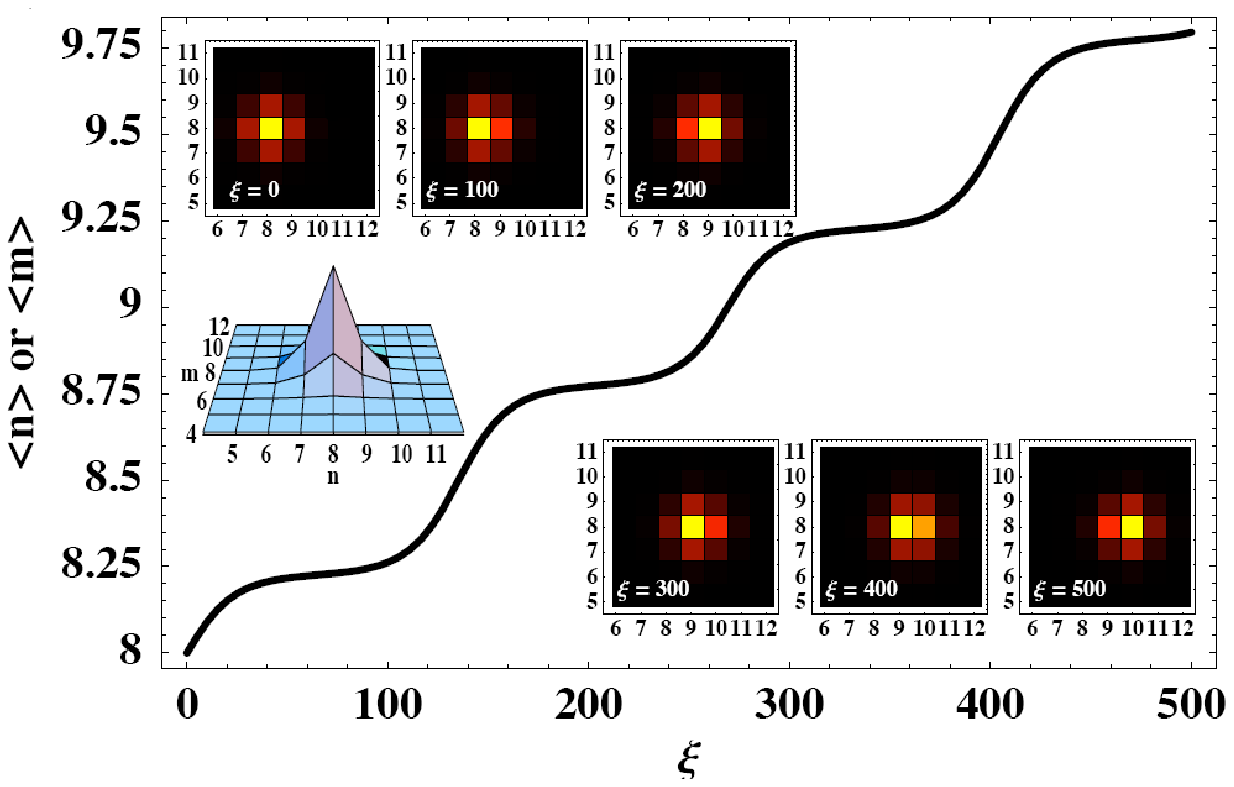}
\caption{Upper figure: PN potential surface of Eq.~(\ref{johansson:eq:sat}) 
for $\gamma=4$ and $P \approx 9.4$. The different stationary solutions 
are marked with arrows.
Adapted from Ref.~\cite{naether2011a}.
Lower figure: The resulting mobility in an axial direction, after applying
a small phase gradient ($k\approx 6 \cdot 10^{-3}$) to a stable 1-site mode, 
which adds an energy just enough to overcome the very small PN barrier 
($\Delta H \approx 2 \cdot 10^{-4}$) to the stationary intermediate 
solution (IS).
Main figure shows motion of center of mass, inset shows profiles at 
different $\xi$. From Ref.~\cite{vicencio2006}. }
\label{fig:johansson-figure05}
\end{center}
\end{figure}
For increasing power, the scenario changes as the first stability exchange 
regime (illustrated in Fig.~\ref{fig:johansson-figure04} for a different 
value of $\gamma$) is reached. 
In 
 Fig.~\ref{fig:johansson-figure05}, two new saddle points corresponding 
to the unstable, asymmetric, stationary 
intermediate solutions (IS) have appeared, 
while the extrema corresponding to the (now stable) 
2-site modes have changed to local minima. Note that the energy landscape 
is almost flat between the simultaneously stable 1-site and 2-site solutions, 
resulting in a very good axial mobility 
(lower plot in Fig.~\ref{fig:johansson-figure05}). Note also how the very 
slowly moving mode in Fig.~\ref{fig:johansson-figure05} clearly traces 
out the local features of the PN potential in the axial directions, where 
the velocity is minimal at each location for the center of mass corresponding 
to IS saddles in the potential surface. 

\begin{figure}[t] \begin{center}
\includegraphics[width=5.5 cm, angle=270]{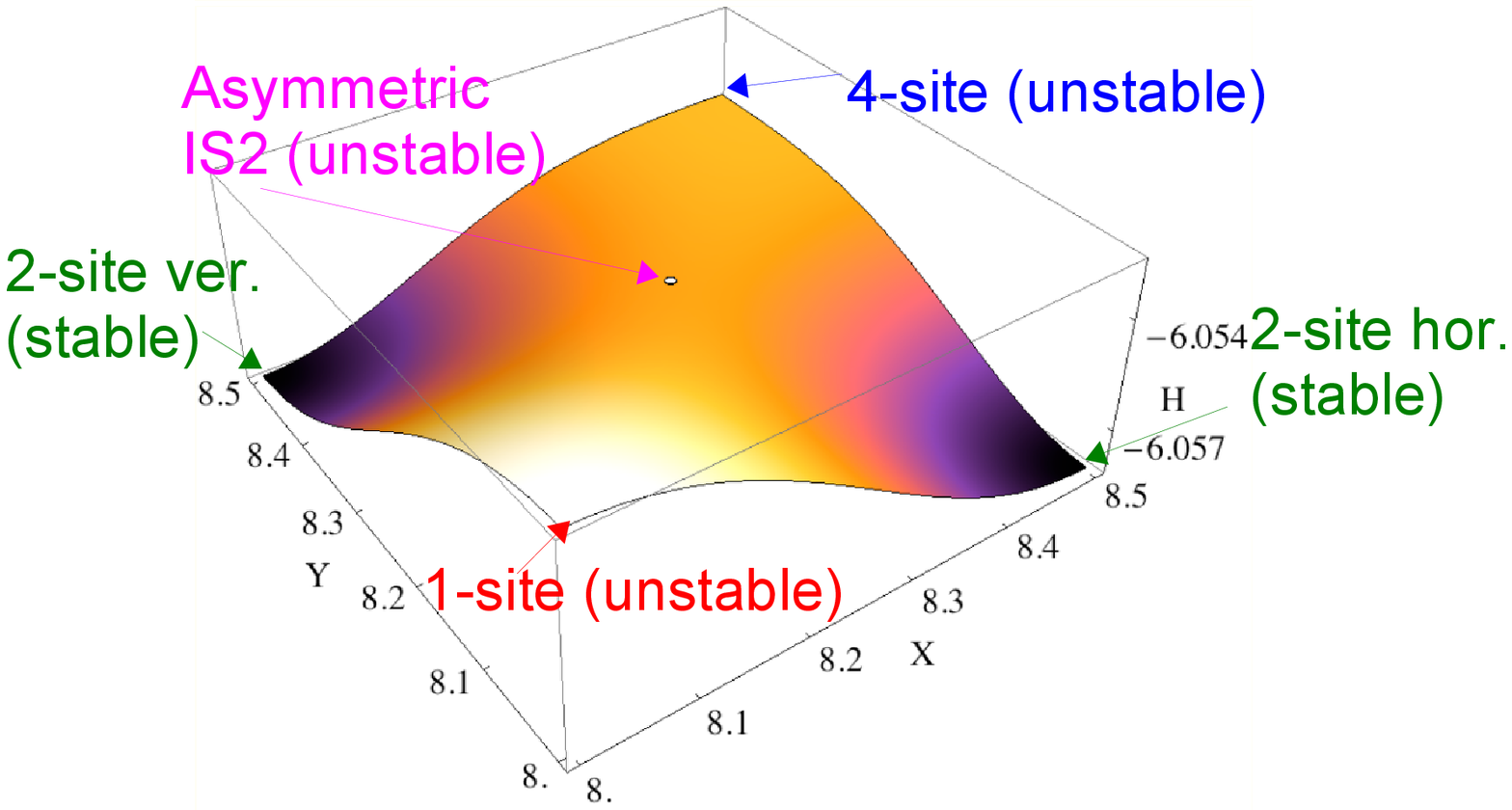}
\includegraphics[width=4.5 cm, angle=270]{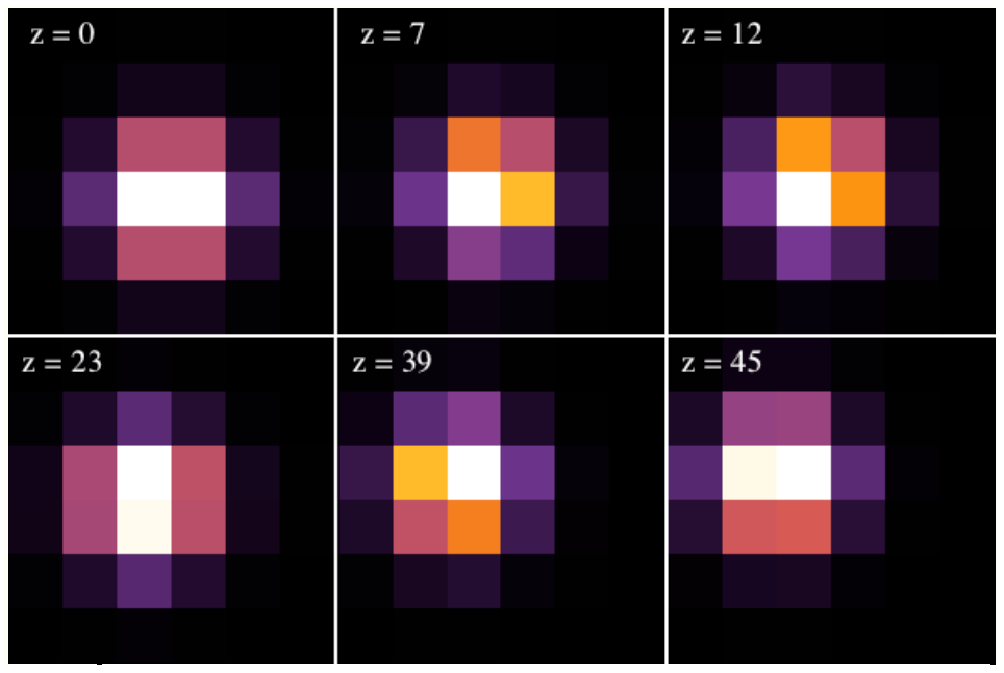}
\caption{Upper figure: PN potential surface of Eq.~(\ref{johansson:eq:sat}) 
for $\gamma=4$ and $P = 10.0$, with stationary solutions 
marked with arrows.
Lower figure: The resulting mobility in a diagonal direction, after applying
a small phase gradient ($|k_x| = |k_y| = 0.018$) to a stable 
horizontal 2-site mode (here, 
$z$ is used instead of $\xi$ to denote the time-like variable in 
Eq.~(\ref{johansson:eq:sat})).
Adapted from Ref.~\cite{naether2011a}.}
\label{fig:johansson-figure06}
\end{center}
\end{figure}
A further increase in $P$ turns the 1-site mode unstable, and the PN 
surface for the regime when only the 2-site modes are stable is shown 
in Fig.~\ref{fig:johansson-figure06}. Note that the topology of the 
surface, with two equivalent minima corresponding to the stable horizontal 
and vertical modes and two local maxima corresponding to the unstable 
1-site and 4-site modes, necessitates a saddle point along the diagonal 
between the maxima, and therefore another asymmetric unstable stationary 
intermediate solution 
(here termed IS2) must exist.  The flatness of the energy landscape 
(note the scale on the $H$-axis) between 
the stable horizontal and vertical 2-site modes implies a new type of mobility 
in the diagonal direction, illustrated in the lower part of 
Fig.~\ref{fig:johansson-figure06}: the soliton moves its center 
along the diagonal by repeatedly transforming between horizontal and vertical 
shapes, passing over the small PN barrier created by the intermediate solution
(see Ref.~\cite{naether2011a} for further illustrations). 

Continuing the increase of power, a fourth regime is reached where 
also the 4-site solution has stabilized (this occurs when the IS2 saddle 
in Fig.~\ref{fig:johansson-figure06} reaches the 4-site max), yielding 
a PN surface with local minima at the stable 2-site and 4-site positions, 
a maximum at the unstable 1-site position, and saddles corresponding 
to new unstable intermediate solutions between 2-site and 4-site 
modes~\cite{naether2011a}. The energy landscape is now almost flat 
 between the 2-site and 4-site positions, resulting again in a very good 
mobility along axial directions but now between the 2-site and 4-site 
modes~\cite{naether2011a}.
 
Finally, a fifth qualitatively different regime is reached when increasing $P$, 
where the 2-site solutions have turned unstable and only the 4-site mode 
is stable~\cite{vicencio2006,naether2011a}. Thus, the PN surface
has only one minimum at the 4-site position, saddles at the 2-site positions 
and maximum at the 1-site position. There are no intermediate solutions 
but, as illustrated in  Ref.~\cite{naether2011a}, the PN potential may 
still be sufficiently smooth and flat to allow for mobility in, e.g., 
diagonal directions with an appropriate initial perturbation. 

A further increase in power yields repeated stability 
exchanges~\cite{vicencio2006,naether2011a}, and so the above described 
five different regimes of qualitatively different PN potentials, and 
their corresponding characteristic mobility properties, will reappear 
repeatedly~\cite{naether2011a}. Among other issues discussed in  
Ref.~\cite{naether2011a}, it was also shown that including weak lattice 
anisotropy breaks the symmetry between the horizontal and vertical 2-site 
modes, thereby allowing for two additional PN surface topologies (see 
Ref.~\cite{naether2011a} for details). In particular, it was seen that 
for a non-negligible anisotropy, all intermediate solutions appear  
on the edges of the surfaces (i.e., scenarios with IS2-type solutions as 
in Fig.~\ref{fig:johansson-figure06} disappear), implying that the best 
mobility for anisotropic lattices should generally appear along lattice 
directions. Thus, in conclusion, calculating the full 2D PN potentials appears 
as a very powerful tool for predicting the directional mobility properties 
in 2D lattices. 



\subsection{The Kagome lattice}
\label{johansson:subsec:kagome} 

In this subsection, we summarize and discuss results obtained 
in Ref.~\cite{vicencio2013} (to which the reader is referred for further 
details and references) regarding the mobility properties of the so called 
``discrete flat-band solitons'' in the 2D Kagome lattice. 
\begin{figure}[t] \begin{center}
\includegraphics[height=8 cm, angle=270]{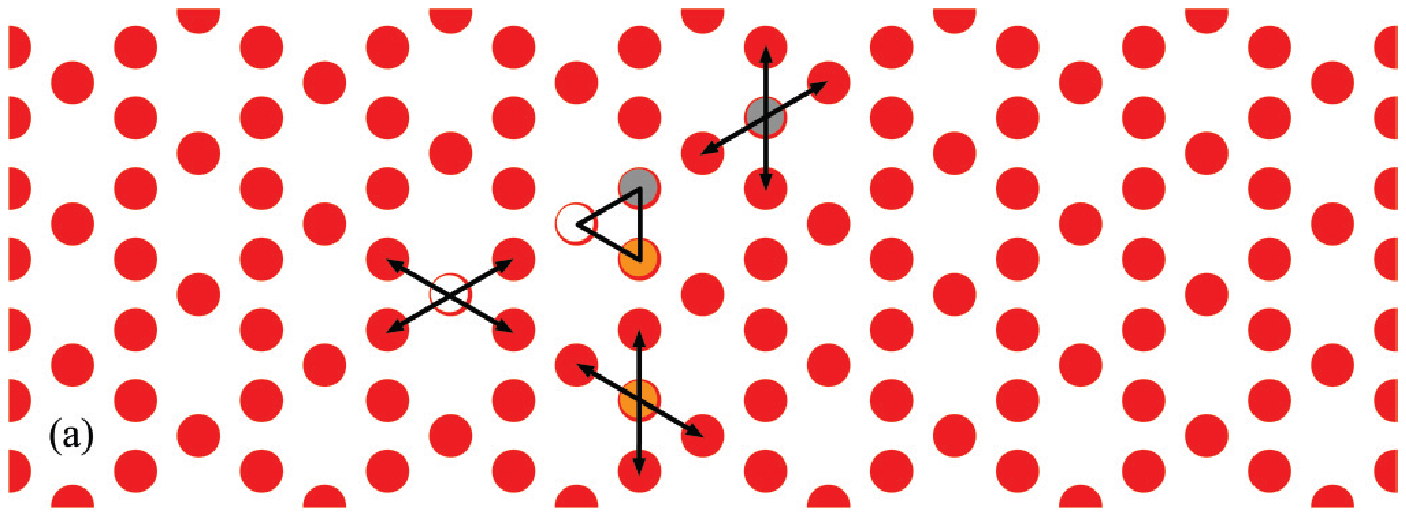}
\includegraphics[width=4.5 cm, angle=270]{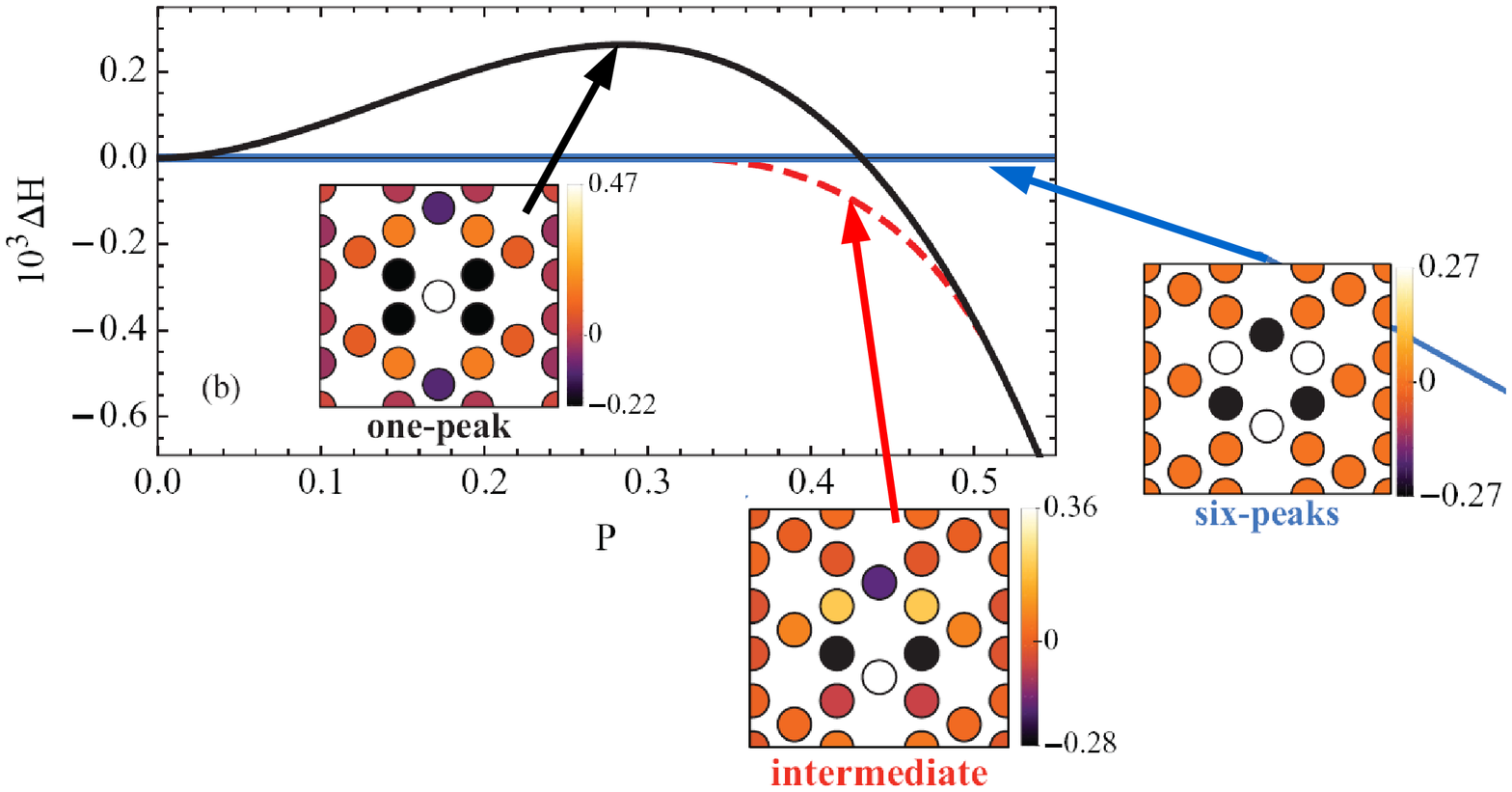}
\caption{Upper figure: Structure of the Kagome lattice showing
a unit cell with three sites (triangle) and the directions of their 
respective nearest-neighbour interactions. 
Lower figure: Hamiltonian (relative to the six-peaks solution) versus 
power for the 
fundamental stationary nonlinear localized modes of 
Eq.~(\ref{johansson:eq:dnls}) with amplitude profiles (for $P=0.43$) 
as indicated. 
Adapted from Ref.~\cite{vicencio2013}.}
\label{fig:johansson-figure07}
\end{center}
\end{figure}
The 
structure of the Kagome lattice is illustrated in 
Fig.~\ref{fig:johansson-figure07}, and as indicated in the figure, it 
can be viewed as a hexagonal lattice with a three-site, triangular unit cell. 
We will consider the ordinary, cubic, on-site DNLS model defined with 
nearest-neighbour interactions according to the Kagome lattice structure 
as indicated in Fig.~\ref{fig:johansson-figure07}, which with the notation 
of Ref.~\cite{vicencio2013} takes the form
\begin{equation}
i\frac{\partial u_{\vec{n}}}{\partial z}+\sum_{\vec{m}} V_{\vec{n},\vec{m}} u_{\vec{m}}
+\gamma |u_{\vec{n}}|^2 u_{\vec{n}} = 0\ ,
\label{johansson:eq:dnls}
\end{equation}
where $z$ corresponds to the time-like variable, 
$u_{\vec{n}}$ represents the field  amplitude at site $\vec{n}$, and 
the sum over $\vec{m}$ is restricted to nearest neighbours to  $\vec{n}$ 
in the
Kagome lattice. Here, it is crucial to note that we consider exclusively 
the case with defocusing nonlinearity, which implies that with a proper 
normalization we can put $\gamma = - V_{\vec{n},\vec{m}} \equiv -1$. 

The linear spectrum ($\gamma = 0$) of Eq.~(\ref{johansson:eq:dnls}) is 
well known (see, e.g., Ref.~\cite{bergman2008}): of its three (connected) 
bands, the lowest one is exactly flat (dispersionless). As shown 
in Ref.~\cite{bergman2008}, the flat band contains as many states as 
the number of closed rings in the lattice, and thus can be considered to 
be built up from ``six-peaks'' (or ``ring'') solutions, where six sites in 
a closed hexagonal 
loop have equal amplitude but alternating phases, with exactly 
zero background. (The zero background results from the frustration property 
of the Kagome lattice:  each site immediately outside a ring mode couples 
identically to two sites in the ring, but since these sites have 
 opposite phases, their contributions cancel out 
due to destructive interference.) The amplitude profile of a six-peaks mode 
is show in the lower, rightmost part of Fig.~\ref{fig:johansson-figure07}. 

For a defocusing nonlinearity ($\gamma < 0$), 
nonlinear stationary solutions to Eq.~(\ref{johansson:eq:dnls}) will bifurcate 
from the lowest-energy linear band, i.e., the flat band~\cite{vicencio2013}. 
It is easily seen, that the single six-peak ring mode is an exact 
(and strictly compact!) stationary solution also in the nonlinear case, and 
that it exists for all possible values of power, $0 \leq P < \infty$.
Consequently, in sharp contrast to the case for ordinary 2D DNLS lattices 
(having dispersive bands) with cubic nonlinearity discussed in the 
beginning of this section, there is no power (norm) threshold for creation 
of localized stationary solutions in the flat-band Kagome lattice.  

Moreover, also other nonlinear stationary solutions bifurcate from linear 
combinations of the degenerate fundamental linear flat-band ring modes, and 
the nonlinearity will generally break the degeneracy of such solutions. 
In contrast to the single-ring, six-peak, solution, these nonlinear 
solutions will generally not remain compact but develop an exponential 
tail~\cite{vicencio2013}, 
as for ``ordinary'' lattice solitons/breathers. Of special interest is 
the mode obtained by adding together two neigbouring ring modes having 
one site in common, which thus in the linear limit gets an amplitude 
twice as large as the other ten sites in the rings. Also this solution 
belongs to a family of nonlinear localized stationary solutions existing 
for all values of power~\cite{vicencio2013}, and in the limit 
$P \rightarrow \infty$ (``anticontinuous limit''), it becomes a single-site 
localized excitation. The profile of this solution, here termed ``one-peak'', 
is illustrated for a small but nonzero power 
in the lower left part of Fig.~\ref{fig:johansson-figure07} 
(note that the two contributing rings are vertically aligned in this figure).

Comparing the Hamiltonian (energy) at fixed power (norm) for these two
families of solutions, it can be checked~\cite{vicencio2013} that the 
single-ring 
(six-peaks) mode has the lowest energy and constitutes the ground state 
of the system  close to the linear limit, while the 
double-ring (one-peak) mode is the ground state for strong nonlinearity. 
Thus, as illustrated in the lower part of Fig.~\ref{fig:johansson-figure07}, 
there is an exchange of stability between these two modes, with a scenario 
similar to what has been described for other models above, with appearance
of an asymmetric, intermediate stationary solution in the exchange regime. 
In fact, the scenario is here analogous to that of 
Fig.~\ref{fig:johansson-figure02} (b), with simultaneous instability of 
the on-site (one-peak) and inter-site (six-peaks) modes and a stable, 
symmetry-broken intermediate solution constituting the ground state of the 
system.  

\begin{figure}[t] \begin{center}
\includegraphics[height=4.5 cm, angle=0]{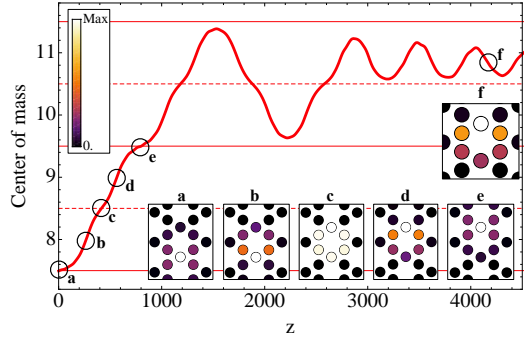}
\caption{Main figure: Evolution of the vertical center of mass when applying 
a very small vertical kick ($k_y=0.009$) to an unstable one-peak solution in 
the stability-exchange regime of Fig.~\ref{fig:johansson-figure07} 
($P = 0.4655$). Horizontal lines mark out the locations of
stationary one-peak (solid) and six-peaks (dashed) modes. 
Insets show intensity profiles $|u_{\vec{n}}(z)|^2$ of the 
travelling mode at the indicated locations (a)-(f). 
From Ref.~\cite{vicencio2013}.}
\label{fig:johansson-figure08}
\end{center}
\end{figure}
Thus, having all previously discussed examples of  connections between 
stability exchange and enhanced mobility in mind, it might not be unexpected 
that good mobility between these strongly localized, fundamental modes may 
appear also here,\footnote{The reader should however be 
cautioned 
that there are counter-examples where mere stability exchange does 
{\em not} imply good mobility, as for the 2D version of 
Eq.~(\ref{johansson:eq:xdnls})~\cite{oster2009}, 
since it does not automatically imply 
that a smooth and flat PN surface exists in the full domain.} 
and at a relatively small power as can be seen from
Fig.~\ref{fig:johansson-figure07}. The 
results from applying a small vertical kick (phase gradient) on an unstable 
one-peak mode in the stability-exchange regime is show in
 Fig.~\ref{fig:johansson-figure08}. As can be seen, the initial movement 
is quite analogous to previously discussed cases (cf., e.g., 
Fig.~\ref{fig:johansson-figure05}) and nicely traces out the features of 
the PN potential in the corresponding direction, with smallest velocities 
in the unstable one-peak ((a), ((e)) and six-peaks ((c)) positions,
and largest velocities in the stable intermediate positions ((b), (d)). 
For the very tiny kick in Fig.~\ref{fig:johansson-figure08}, the mode 
quickly loses its surplus energy due to radiation effects, and finally 
gets trapped with small oscillations around an intermediate 
stationary solution ((f)), 
constituting its symmetry-broken ground state in this
regime. As discussed further in Ref.~\cite{vicencio2013}, the distance
travelled in the lattice may be controlled to some extent by the kick strength, 
thus yielding a mechanism for controlled transfer, in particular 
directions of the Kagome lattice, of small-power strongly localized modes 
in a 2D DNLS-lattice with standard (cubic) nonlinearity. 

We end this subsection with a brief mentioning of some earlier works 
discussing nonlinear localized modes in Kagome lattices. Law 
et al.~\cite{law2009} also considered the defocusing case, but concentrated 
on vortices and complex structures mainly in the strong-nonlinearity regime, 
without making connections to flat-band linear modes or mobility. 
Zhu et al.~\cite{zhu2010} studied defect solitons with saturable nonlinearity, 
and Molina~\cite{molina2012} localized modes in nonlinear photonic nanoribbons; 
however, both these works considered the case of focusing nonlinearity, which 
follows the standard 2D NLS phenomenology with threshold etc., since the 
upper band is non-degenerate~\cite{bergman2008,vicencio2013}.

\section{Travelling discrete dissipative solitons with intrinsic gain}
\label{johansson:sec:gain}

The discussion in the previous sections has dealt exclusively with conservative
lattices (i.e., conserved energy), and in addition we have seen that the 
analysis of breather mobility in terms of PN potentials needs a second 
quantity to be (at least approximately) conserved (typically action, 
or norm/power for DNLS-type models). As we also discussed, unless we succeed 
to tune our model parameters into an exact ``transparent point'', and succeed 
to give our breather a ``special'' velocity (or succeed to find some other 
exceptional system like an integrable model), moving breathers in
Hamiltonian lattices  are 
not exponentially localized outside their main core, but develop an extended 
tail due to radiation even when the PN potentials are very smooth and flat. 
The tails may be very weak 
(as e.g. the example shown in Fig.~\ref{fig:johansson-figure03} (b)), 
but due to the radiation continuously emitted, a breather travelling 
in a large lattice will typically in the end get trapped around some minumum 
of the PN potential. 

In a dissipative environment, the situation will naturally be quite 
different. Pure losses will evidently damp out the radiative tails, but also 
the energy of the breather core. However, if there is some additional 
intrinsic gain 
mechanism (such as for a lasing system in optics), one could hope to, under 
certain conditions, establish a balance (at least when averaged over time) 
where the gain is strong enough to support a (possibly strongly localized) 
moving breather indefinitely, 
but weak enough not to destroy the exponentially decaying tail. That this 
indeed is possible, under certain conditions, was demonstrated recently 
in Ref.~\cite{johansson2014}
for a model of a 1D waveguide array in an active Kerr medium with 
intrinsic, saturable gain and damping. Here we will briefly summarize and 
discuss some of the main results from Ref.~\cite{johansson2014}, to 
which we refer for details and further references. 

Under the assumption of a pure on-site Kerr nonlinearity, the model studied 
in Ref.~\cite{johansson2014} (originally suggested by Rozanov's group, 
see Ref.~\cite{kiselev2008} and references therein) 
is a generalized Discrete Ginzburg-Landau (DGL) type model which 
can be written in the form
%
\begin{equation}
i \dot \psi_n + C(\psi_{n-1}+\psi_{n+1}) 
+\left( V_n + |\psi_n|^2  - i f_d(|\psi_n|^2) \right) \psi_n =0 .
\label{johansson:eq:master-eq}
\end{equation}
Thus, Eq.~(\ref{johansson:eq:master-eq}) 
is equivalent to the pure on-site version of 
Eq.~(\ref{johansson:eq:xdnls}) (we will comment briefly below also 
on the extension to intersite nonlinearities, $K_4 = K_5 \neq 0$, which 
was discussed in some detail in Ref.~\cite{johansson2014}), with the addition 
of a possible linear (real) on-site potential $V_n$ ($V_n \equiv 0$ for 
periodic lattice), and, most importantly, a (real) function 
$f_d (|\psi_n|^2)$ describing the amplification and absorption characteristics 
of each waveguide. \footnote{For simplicity, $C$ is chosen real, i.e., 
absorption 
or gain in the medium between the waveguides is neglected.} As in 
Ref.~\cite{kiselev2008} (and references therein), the function 
$f_d (x)$ is chosen to include linear and saturable absorption, as well 
as saturable gain, and can after proper normalizations be taken on the 
four-parameter form
\begin{equation}
f_d(x) = - \delta + \frac{g}{1+x}  - \frac{a}{1+bx} ,\quad \delta, g, a, b >0 . 
\label{johansson:eq:fd}
\end{equation}
The parameters describe, respectively, linear losses ($\delta$), saturable 
gain strength ($g$), saturable absorption strength ($a$), and ratio
between gain and absorption saturation intensities ($b$). As detailed 
in Refs.~\cite{kiselev2008,johansson2014}, the conditions to have
localized modes which simultaneously should have a stable zero-amplitude tail, 
and a core with a non-zero, non-decaying amplitude, put several restrictions 
on the possible parameter intervals (it also follows that $b > 1$, i.e., 
the gain must saturate at a higher intensity than the damping). The observant 
reader will notice that expanding Eq.~(\ref{johansson:eq:fd}) to second 
order in $x$ yields a cubic-quintic DGL model, which may be a more familiar 
system (see, e.g., Ref.~\cite{efremides2003}). However, as was found 
empirically by extensive numerical searches in Ref.~\cite{johansson2014}, 
the relevant solutions describing moving localized modes essentially result 
from the strong saturabilities of the gain and damping parts on different 
intensity scales, and therefore in regimes not well described by
a cubic-quintic approximation. 

\begin{figure}[t] \begin{center}
\includegraphics[height=5.81 cm, angle=270]{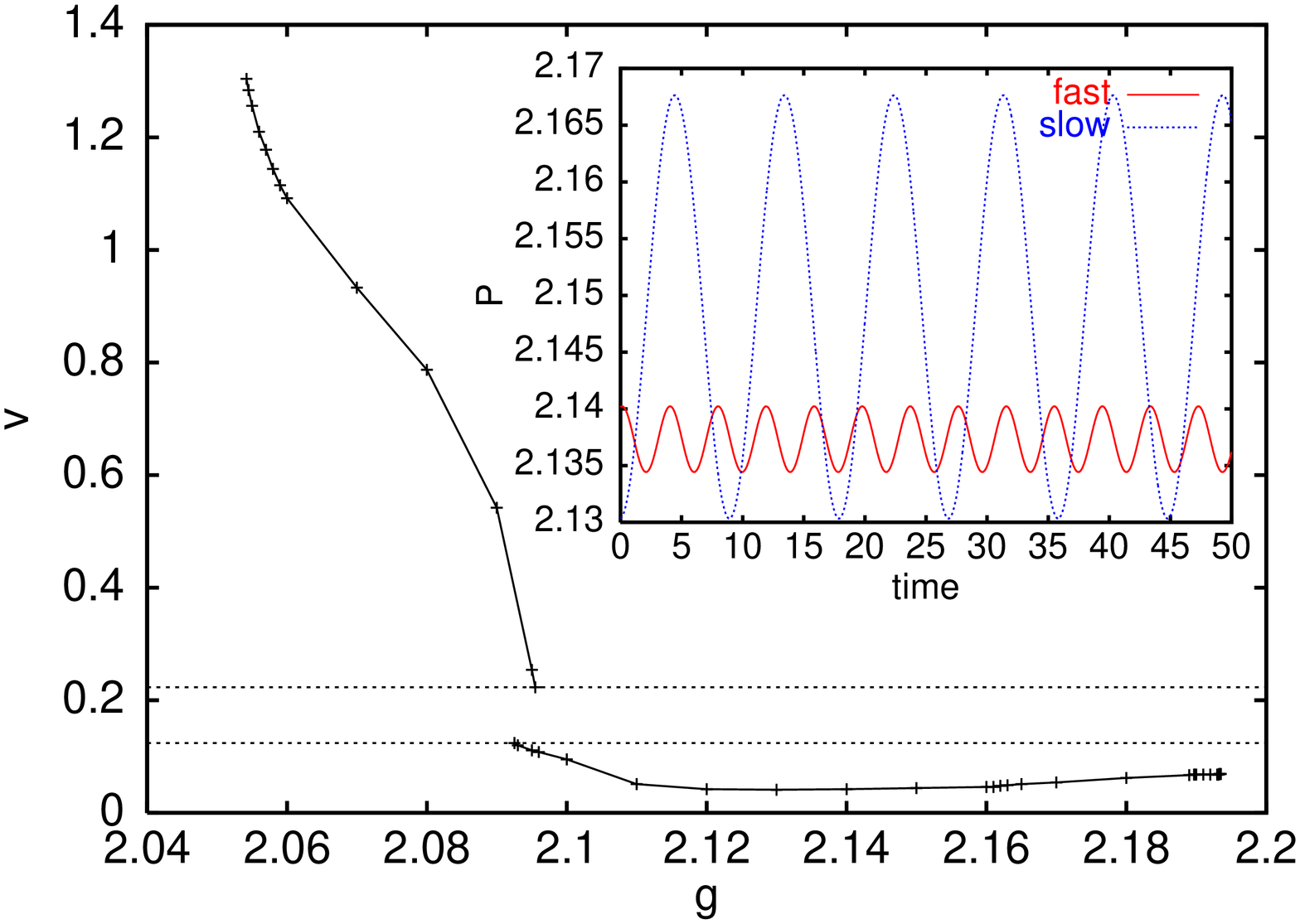}
\includegraphics[height=5.81 cm, angle=270]{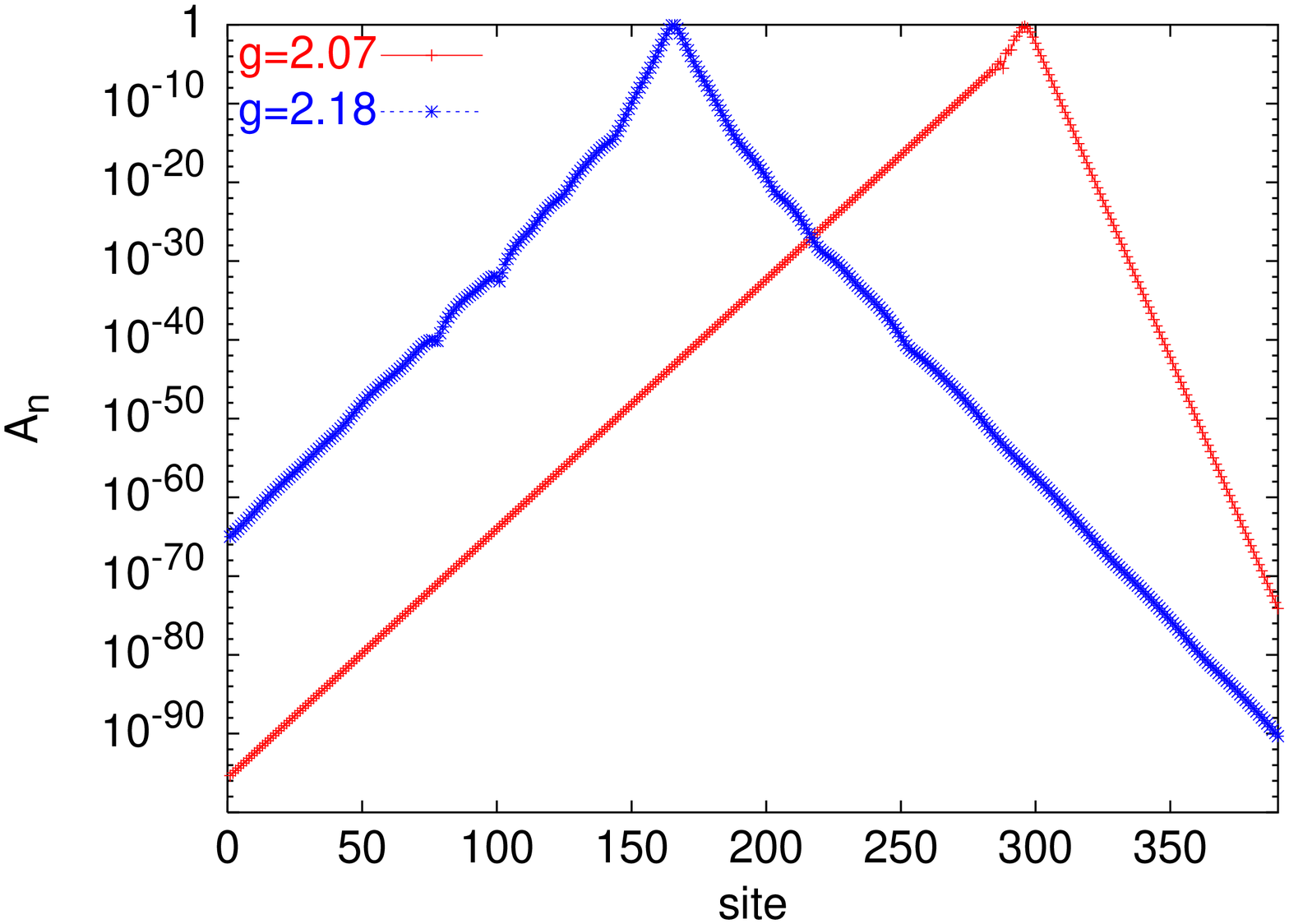}
\caption{Left figure: Velocity $v$ versus gain parameter $g$ for moving 
localized solutions of Eq.~(\ref{johansson:eq:master-eq}) with 
$V_n \equiv 0$, $C=1$, and
parameters in Eq.~(\ref{johansson:eq:fd}) chosen as $\delta = 1$, $a=2$, 
$b=10$. Horizontal lines indicate a gap of ``forbidden'' velocities, 
and inset shows norm oscillations for two bistable solutions at 
$g=2.095$: a fast solution with small oscillations and a slow solution with 
large oscillations. Right figure: Snap shops of intensity $A_n = |\psi_n|^2$
for two right-moving solutions with $g=2.07$ (right peak, fast mode) and 
$g=2.18$ (left peak, slow mode).
From Ref.~\cite{johansson2014}.}
\label{fig:johansson-figure09}
\end{center}
\end{figure}
In Fig.~\ref{fig:johansson-figure09} we illustrate a  typical scenario 
with moving, strongly localized solutions for a ``suitable'' regime of 
parameter values (see Ref.~\cite{johansson2014} for further discussions 
on the influence of parameter variations). As can be seen, 
gain-driven, travelling 
discrete solitons exist as exact exponentially localized solutions at 
specific velocities, although in a rather narrow interval for the gain 
parameter. It is important to note that, in contrast to conservative systems 
where there are continuous families of breathers/solitons as discussed above 
(which can be parametrized e.g. using the norm/action, or frequency, as 
parameter), breathers/solitons in dissipative systems generically appear 
as isolated attractors where an appropriate balance between energy 
input and dissipation can be established
(see, e.g., Refs.~\cite{flach2008a,flach2008b} and references therein). Here, 
for most values of $g$ where moving solitons are found as attractors, 
they have a well-defined, single velocity $v$, which typically increases for 
smaller gain (although the dependence generally is not strictly monotonous 
as seen in Fig.~\ref{fig:johansson-figure09}, left part). Note also
the division in a ``fast'' and a ``slow'' branch, with a forbidden velocity 
gap and a small regime of bistability. 

The exponential localization of the moving solitons is illustrated in 
the right part of Fig.~\ref{fig:johansson-figure09}. Two features are 
noteworthy: (i) a crossover between one decay rate around the soliton core, 
and another (generally weaker) in the tails; (ii) the stronger decay rate 
in the forward than in the backward direction (particularly visible 
for the fast soliton). As discussed in Ref.~\cite{marin2001}, the 
latter is an effect of the radiation emitted from the breather core during 
its motion being Doppler shifted. 

As can be seen from the inset in the left figure in 
Fig.~\ref{fig:johansson-figure09}, the norm (power) of the moving solutions is 
not constant but oscillates time-periodically during the motion; similar 
oscillations (with the same period) occur for the Hamiltonian 
(energy)~\cite{johansson2014}). The necessity for such oscillations in order 
to sustain an exact moving, strongly localized discrete soliton has a 
simple, intuitive 
interpretation in terms of the PN potential of the corresponding conservative 
system: in order to overcome the PN barrier and travel with a constant 
average velocity, the soliton may adjust its 
internal degrees of freedom to its lattice position by locally absorbing and 
emitting ``suitable'' amounts of norm and Hamiltonian via the gain and 
damping terms, respectively. As seen in Fig.~\ref{fig:johansson-figure09}, 
the largest oscillations typically appear for the ``slow'' solitons appearing 
for the larger gain values; essentially these solitons also have a higher 
peak power and are more strongly localized, and therefore the corresponding 
effective PN potential should be stronger.

In the main part of the regime where moving solitons exist, the 
oscillations in $P$ and $H$ are 1:1-locked with the soliton translation, 
i.e., the soliton returns to its initial shape after translation with one site. 
However, as discussed in more detail in Ref.~\cite{johansson2014}, 
when approaching
the rightmost part of the existence regime for $g$ in 
Fig.~\ref{fig:johansson-figure09}, the soliton undergoes a sequence 
of period-doublings (i.e., the soliton does not return to its initial 
shape until after a translation with $2^k$ sites), until it loses its 
regular movement and enters a regime of apparently random motion. For other 
parameter values, also small windows of period-3 translational 
motion were found in Ref.~\cite{johansson2014}.

As mentioned above, the existence regime for moving discrete solitons 
in the model (\ref{johansson:eq:master-eq}) is quite narrow, which can be 
related to the non-negligible PN barrier for strongly localized modes of 
the ordinary (conservative) cubic DNLS model. However, as was discussed in 
Sec.~\ref{johansson:sec:1D}, inter-site nonlinearities as 
in Eq.~(\ref{johansson:eq:xdnls}) may drastically 
decrease the PN potential and improve the mobility. One may therefore
suspect that, similarly, inclusion of inter-site nonlinearities  also may 
increase the existence regime for moving solitons in the gain-damped DGL 
model. Without going into details (see Ref.~\cite{johansson2014}), the answer 
is strongly in the affirmative. For example, for the same parameter values
as in Fig.~\ref{fig:johansson-figure09}, the existence regime in $g$ is 
about five times larger when $K_4=K_5=-0.2$. Another effect of the weakened 
PN barrier is, that the internal oscillation of the soliton may 
become decoupled 
from its translational motion, resulting in a quasiperiodically moving 
soliton which, although it moves with constant velocity, never exactly 
returns to its initial shape in the lattice~\cite{johansson2014}. Also 
tiny regimes of non-trivial phase locking (e.g., makes two internal 
oscillations while moving five sites) were observed in 
Ref.~\cite{johansson2014}. 

\begin{figure}[t] \begin{center}
\includegraphics[height=5.81 cm, angle=270]{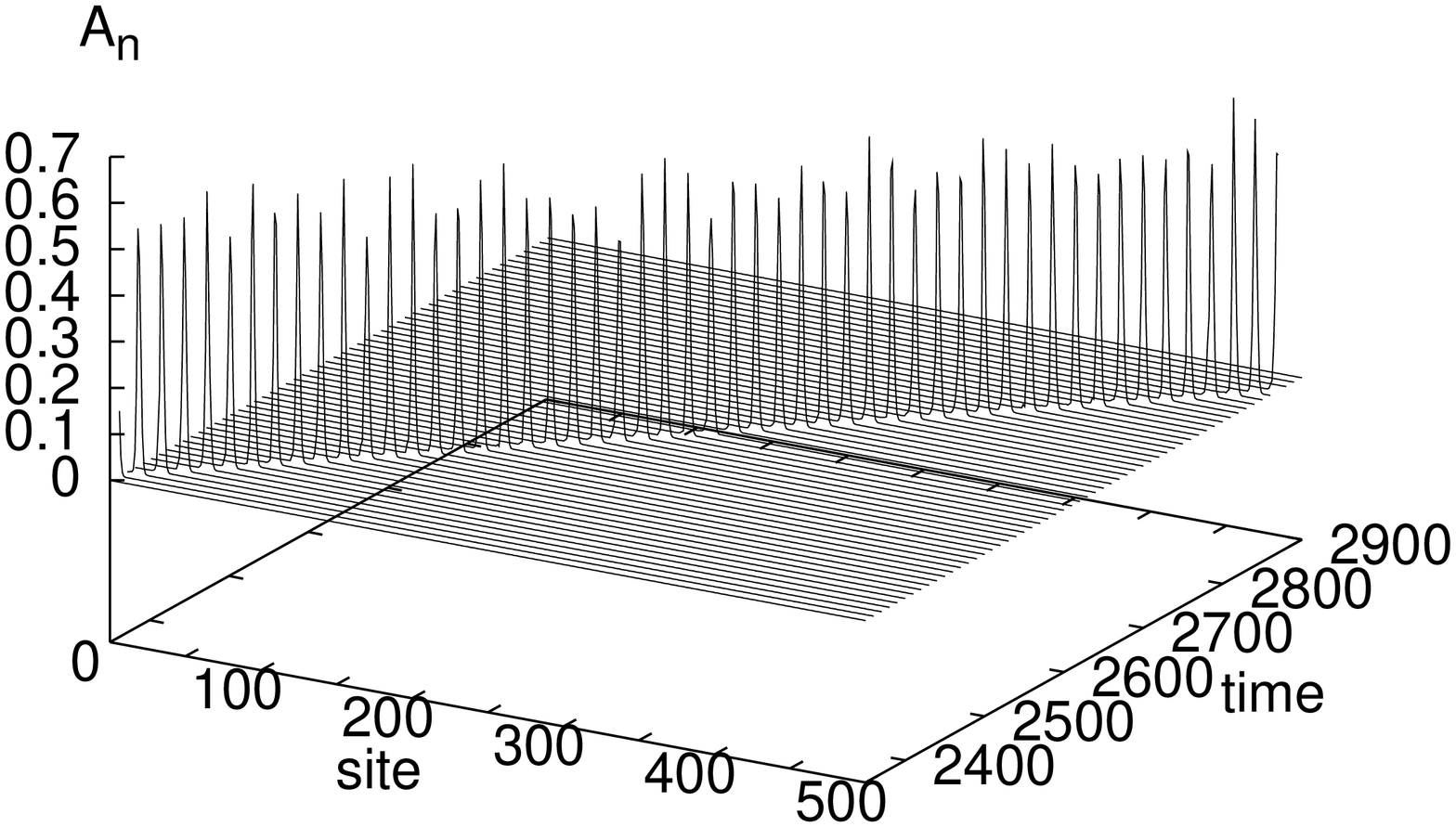}
\includegraphics[height=5.81 cm, angle=270]{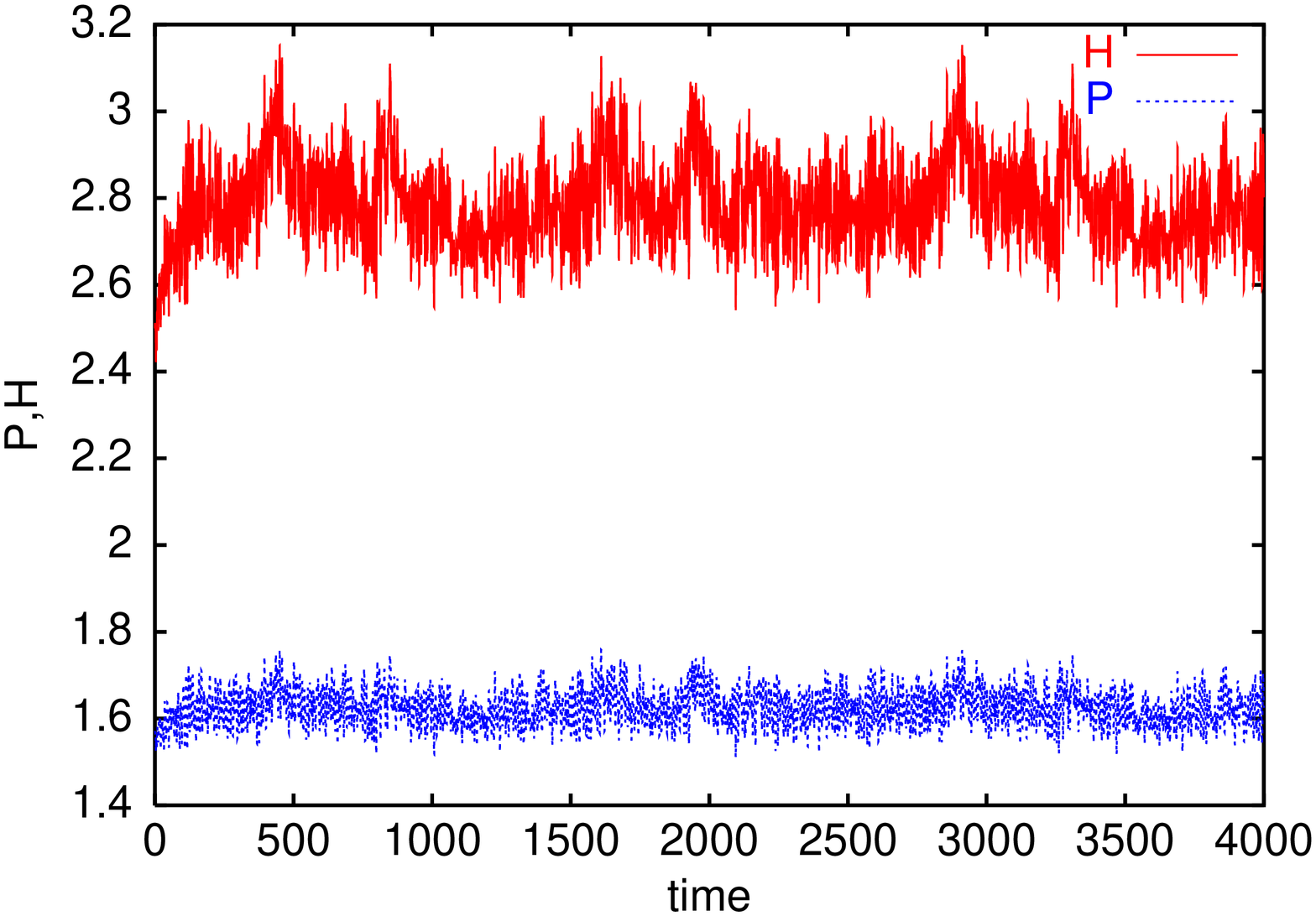}
\caption{Left figure: Intensity distribution for soliton moving with 
constant velocity $v\approx 0.977$ in a lattice with a uniformly distributed, 
disordered on-site potential $V_n \in [-0.1,0.1]$. $g=2.06$, other 
parameters same as in Fig.~\ref{fig:johansson-figure09}. 
Right figure: The corresponding oscillations for the Hamiltonian (upper) and 
norm (lower).
From Ref.~\cite{johansson2014}.}
\label{fig:johansson-figure10}
\end{center}
\end{figure}
As a final illustration of the ability of the moving discrete
dissipative solitons to keep 
moving with constant velocity by adjusting to 
their local environment in the lattice, 
we show in Fig.~\ref{fig:johansson-figure10}
an example from Ref.~\cite{johansson2014} 
of a moving soliton in a weakly {\em disordered} lattice (again 
with pure on-site nonlinearity, $K_4=K_5=0$). 
Although the on-site potential $V_n$ is chosen randomly from a uniform 
distribution, the soliton moves indefinitely (here in a lattice with 2405 
sites and periodic boundary conditions~\cite{johansson2014}) with constant 
velocity! It does so by, at each point, adjusting its internal parameters
according to its local environment. As illustrated in the right part 
of Fig.~\ref{fig:johansson-figure10}, this results in irregular oscillations 
of norm and Hamiltonian, compensating for the irregularities in the lattice.
A careful look at these curves confirms this scenario: although they may 
look random, in fact they are not. After an initial transient, they 
periodically repeat themselves {\em exactly} with a period of 2462 time units, 
corresponding to one round trip in the lattice!

Let us end this section with some brief discussion about other related 
works (a more extensive discussion was given in Ref.~\cite{johansson2014}).
Surely this is not the first observation of moving discrete 
dissipative solitons/breathers;
see, e.g., the reviews~\cite{flach2008a,flach2008b}. 
Indeed, many aspects of the mobility scenarios described here are analogous 
to what has been reported earlier for other systems: The existence 
of two types of ``fast'' and ``slow'' breathers with exponentially decaying
phonon 
tails were described for the damped-driven FK model in Ref.~\cite{marin2001},
similar modes have been discussed in the context of ``discrete cavity 
solitons'' in optics 
(see, e.g., Refs.~\cite{egorov2007,yulin2011,egorov2013}), and 
also experimentally moving localized modes with similar properties have 
been observed in damped-driven electrical 
lattices~\cite{english2008,english2010,english2013}. However, conceptually, 
all these systems are different from the model discussed here and 
in Ref.~\cite{johansson2014}, in the sense that they require an explicit, 
uniform {\em external} driving to supply the necessary energy to 
compensate for the damping. As a consequence, the moving breathers in these 
systems are not strictly localized but decay (exponentially) 
towards a tail of constant, non-zero amplitude, implying that their energy 
would increase towards infinity for increasing system size. By contrast, 
in Eq.~(\ref{johansson:eq:master-eq}) the gain results from purely 
{\em intrinsic} properies of the medium where the soliton propagates
(such as, e.g., a lasing system), allowing for propagating 
finite-energy solitons 
with tails decaying towards zero.

We considered here only one particular model of a gain/damped system 
(admittedly, rather special 
with many different ingredients). It would of course be highly interesting 
to investigate whether travelling localized modes, 
driven by some intrinsic gain 
mechanism, can exist also in more general physical lattice systems. One 
particularly interesting issue, pointed out to us by Mike Russell, is 
the suggestion that the quodons in mica-like systems could travel for 
macroscopic distances thanks to an intrinsic gain mechanism, resulting 
from the lattice being in a metastable configuration~\cite{russell2014}.

\section{Mobility of quantum lattice compactons}
\label{johansson:sec:QLC}

So far, we only discussed mobility of nonlinear lattice excitations using 
the language of classical physics. However, in 
many applications of the discrete breather/soliton concept, quantum 
mechanical effects may be important, and therefore it is highly relevant to 
investigate to what extent the above described scenarios for mobility 
survive under quantum fluctuations. The literature on 
``quantum discrete breathers'' is huge and we do not make any attempt 
to give a complete review of this topic here, but refer the reader to 
Refs.~\cite{flach2008a} and  \cite{pinto2011} for discussions and further 
references. Let us just recapitulate some basic facts. Quantum mechanics, in 
the language of a many-body Schr{\"o}dinger equation, is linear, and for 
a periodic lattice the Hamiltonian is invariant under lattice 
translations. Thus, all eigenstates must obey the Bloch theorem, meaning
that they are necessarily delocalized and the probability of finding a 
particular number of excitation quanta (``particles'') at a certain site 
must be the same at \emph{any} site. However, it is possible to define 
localization in another sense, looking instead at \emph{correlations}. 
Following Eilbeck~\cite{eilbeck2003b}, 
a quantum analogue of a classical localized breather may 
then be defined as an eigenstate with a high probability of having 
\emph{many} quanta localized on the \emph{same} site (or, more generally, 
identified as many-particle bound states with correlation functions 
exponentially decaying in space~\cite{wang1996}). 

Alternatively, if we insist 
on creating a quantum state which, like a classical soliton/breather, is 
localized at some \emph{specific} site(s), we need to take an appropriate 
superposition of eigenstates. As discussed in 
 Refs.~\cite{flach2008a,pinto2011,eilbeck2003b} 
(and references therein), it is expected 
that when a classical nonlinear Hamiltonian lattice possesses exact 
discrete breathers/solitons, its quantum counterpart contains nearly 
degenerate bands of eigenstates, corresponding to specific 
many-particle 
bound states with different crystal momenta (``breather bands''). 
The bandwith of such a band is then proportional to the inverse of a 
``tunneling time'', describing the time it takes for a semiclassical breather
to perform a quantum tunneling from one site to the next. The tunneling time 
should become infinite in the classical limit. Note that this quantum breather 
tunneling is a purely quantum effect of a very different nature than the 
coherent mobility of a classical breather. Even though, if the breather band 
is well isolated from other bands, a localized excitation created from a 
superposition of its eigenstates will remain localized in terms of 
correlations (probability to find many particles at the same site remains 
large), it will spread symmetrically in the lattice in terms of the 
expectation value of the local excitation number operator. See e.g. 
Ref.~\cite{falvo2006} for explicit illustrations of this scenario. 

Although quite much effort has been spent on understanding 
various properties of quantum discrete breathers~\cite{flach2008a,pinto2011}, 
to the best of our knowledge very little has been known about the 
quantum counterparts to the classically {\em moving} breathers, and 
in particular whether the concepts of PN potential and barrier have any 
relevance when quantum effects become strong. Clearly, a necessary condition 
for these concepts to make sense must be that the breather band is sufficiently 
narrow for the quantum tunneling time to be much larger than the inverse 
classical velocity (i.e., the time it takes for the classical breather to 
move one lattice site); otherwise, the probability distribution for the 
quantum breather will spread through
tunneling before 
its center has had the time to perform a translation in a given direction. 
Thus, the approach 
of tracing out a PN potential by imagining an infinitely slow breather 
movement makes sense only in the classical limit. 

In Ref.~\cite{jason2013} we addressed some of these issues in the 
context of a 1D \emph{extended Bose-Hubbard} (eBH) model, which is 
a quantum version of the classical extended DNLS model 
(\ref{johansson:eq:xdnls}), for which the quantum Hamiltonian can be 
written in the form \cite{jason2012,jason2013}
\begin{eqnarray}\label{johansson:eq:eBHM}
\hat{H}_{eBH} = 
\sum_{i=1}^f \left\{ 
\frac{1}{2}Q_1 \hat{N}_i + Q_2 \hat{a}_{i+1}^{\dagger} \hat{a}_{i} 
+\frac{1}{2}Q_3\hat{N}_{i}^{2} 
+ Q_4 \left [
2 \hat{N}_{i}\hat{N}_{i+1}+(\hat{a}_{i+1}^{\dagger})^2(\hat{a}_i)^2 \right]
\right.
\nonumber \\ 
\left.
+
2 Q_5\left[ (\hat{a}_{i}^{\dagger})^2 + (\hat{a}_{i+1}^{\dagger})^2 \right]
\hat{a}_i \hat{a}_{i+1} 
\right\} + \textrm{H.c.} .
\end{eqnarray}  
Here $f$ is the number of sites, 
$\hat{a}_{i}^{\dagger} (\hat{a}_{i})$ is the bosonic creation 
(annihilation) operator, and $\hat{N}_i = \hat{a}_{i}^{\dagger}\hat{a}_{i}$ the 
corresponding number operator for particles at site $i$ (H.c.\ is Hermitian 
conjugate). The total number of particles $N$ is conserved since the 
total number operator $\hat{N}=\sum_i\hat{N}_i$ commutes with $\hat{H}_{eBH}$. 
This model appears e.g. in the study of ultracold bosonic atoms
in optical lattices; see the very recent review \cite{dutta2014} 
for extensive discussions and further references (a shorter introduction with 
some additional references was also given recently in Ref.~\cite{jason2014}). 
When $Q_4=Q_5=0$, this is just the ordinary (on-site) Bose-Hubbard model, 
which is a standard model for cold atoms in optical lattices~\cite{dutta2014} 
and also widely studied in the field of quantum breathers since it is 
the quantum counterpart of the ordinary DNLS 
model~\cite{flach2008a,pinto2011,eilbeck2003b}. 
Physically, $Q_2$ represents single-particle tunneling between neighboring 
sites and $Q_3$ a local (on-site) two-body interaction ($Q_1$ defines the 
single-particle energy scale). Also the additional 
nearest-neighbour interaction terms have 
simple physical interpretations: the first $Q_4$-term describes a 
density-density interaction between neighboring sites, the second a 
coherent tunneling of a particle pair, while the  $Q_5$-terms describe 
density-dependent tunnelings since they depend on the number of particles 
at the site the particle is tunneling to and from, 
respectively~\cite{dutta2014}. Taking the classical limit, 
$N\rightarrow \infty$, in an appropriate way~\cite{jason2013} 
results in the Hamiltonian (\ref{johansson:eq:xdnlsham}) for a 
normalized classical 
field $\Psi_i$ with $P=1$ and $|\Psi_i|^2 = <\hat{N}_i> / N$,  after 
a gauge transformation removing $Q_1$, a rescaling putting $Q_3 = -1/2N$, 
and parameter identifications $K_2=Q_2$, $K_4=Q_4/N$ and $K_5=Q_5/N$. 

As is well known~\cite{eilbeck2003b,flach2008a,pinto2011}, computational 
limitations are generally putting severe restrictions on the abilities 
to study quantum properties of classical discrete breathers with exact 
diagonalization, and this is most certainly so also concerning mobility issues. 
Ideally, we would like to study systems with 
many particles to get in contact with the classical world, and large lattices 
to observe localization and translation over some distances. However, 
the dimension of the matrices obtained from 
Eq.~(\ref{johansson:eq:eBHM}) for a given $N$ grows as $(N+f-1)!/N!(f-1)!$, 
so if we wish to study large lattices we are restricted to very few particles, 
and if we wish to study many particles we are restricted to very few sites! 
Here, we may use the latter approach due to a special property for discrete
soliton solutions of the classical model $(\ref{johansson:eq:xdnls})$: as 
was found in Ref.~\cite{oster2003}, at specific parameter values the 
solitons become 
strictly \emph{compact}, i.e., completely localized at a small 
(in fact, arbitrary) 
number of sites with exact zero amplitude outside. Of particular interest 
here is the symmetric inter-site breather denoted (+,+) in 
Fig.~\ref{fig:johansson-figure02}, which compactifies into a two-site 
compacton when $K_5=-K_2/P$, 
where the effective tunneling to outside neighboring sites vanishes. 
In fact, this is precisely the case illustrated in 
Fig.~\ref{fig:johansson-figure02}(b) as also indicated by the (+,+) 
profile in the inset. Thus, for some interval in $K_4$ close to the 
bifurcation points, extremely narrow mobile classical 
solutions exist as was also 
confirmed by direct numerical integrations in Refs.~\cite{oster2003} and 
\cite{jason2013}
(if $K_5=-K_2/P$ exactly, the core of the 
moving classical state will have a rapidly decreasing 
exponential tail which compactifies each 
time it passes an intersite configuration; if the condition is not exactly 
fulfilled its core always decays exponentially but very rapidly as illustrated 
in Fig.~\ref{fig:johansson-figure03}(b)).

Thus, by focusing on quantum counterparts to the compact classical modes, 
we may restrict our studies to very small lattices in order to investigate 
their mobiliity; here we discuss results obtained in Ref.~\cite{jason2013}
for $f=4$ and periodic boundary conditions. For the quantum model
(\ref{johansson:eq:eBHM}), it can be shown that only one-site classical 
compactons have counterparts which are exactly compact also as quantum 
eigenstates~\cite{jason2012,jason2013}. However, the one-site compactons are 
less interesting in the present context since they are not classically mobile. 
The
two-site compactons correspond instead to quantum states with a small, 
and in the classical limit vanishing, probability of finding particles 
spread out over more than two sites~\cite{jason2012}. 
In the neighbourhood of a classical stability exchange region 
as in Fig.~\ref{fig:johansson-figure02} (b), the mobile two-site 
compacton also becomes the ground state when $K_4$ is decreased. 
Thus, well localized quantum 
states may be constructed by taking properly chosen~\cite{jason2013} 
linear combinations of eigenstates in the lowest-energy band.
If left untouched, these states will spread through tunneling with 
a tunneling time increasing with $N$ as discussed above. Far from 
the classical stability exchange regime, where 
the ground-state band is narrow and well isolated from 
other bands, the tunneling times are large and grow rapidly (exponentially) 
with $N$; 
however, approaching the classical stability exchange several bands will 
interact and/or cross, resulting in tunneling times becoming much shorter 
and only slowly increasing with $N$~\cite{jason2013}. An analogous rapid
spreading resulting from hybridization of bands having their main particle 
occupation on a single site and on two sites, respectively, was also 
briefly mentioned in Ref.~\cite{dorignac2004} for a few-particle Bose-Hubbard 
system extended with 
three-particle on-site repulsion (corresponding in the classical limit to a 
cubic-quintic on-site DNLS equation with competing nonlinearities), and was 
described in more detail in Ref.~\cite{falvo2006} 
for another extended Bose-Hubbard model, 
with on-site and pure density-density interactions between neighbouring sites
(i.e., keeping only the first of the $Q_4$-terms in 
Eq.~(\ref{johansson:eq:eBHM})).  

\begin{figure}[t] \begin{center}
\includegraphics[width=10 cm, angle=0]{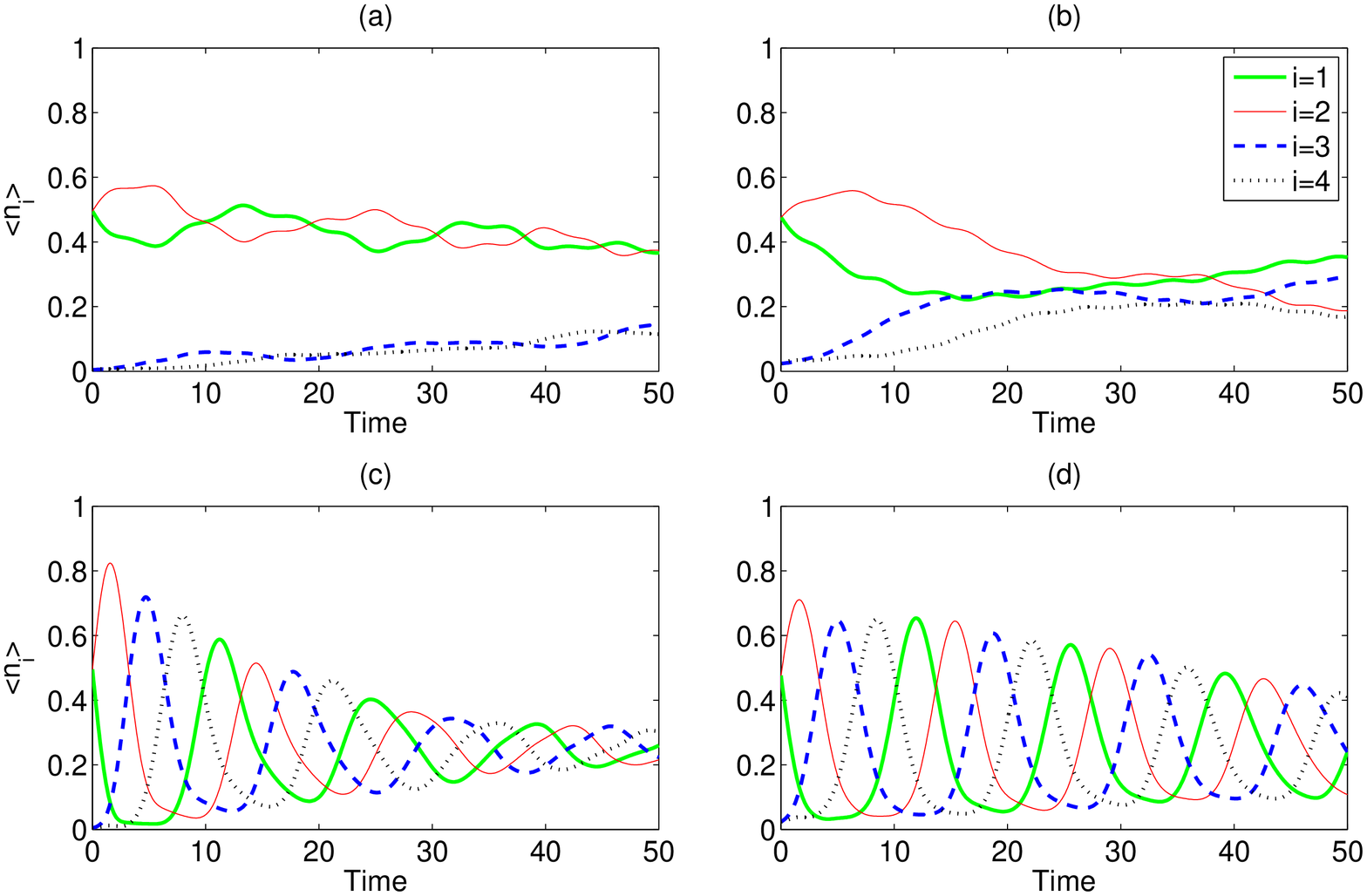}
\includegraphics[width=4 cm, angle=0]{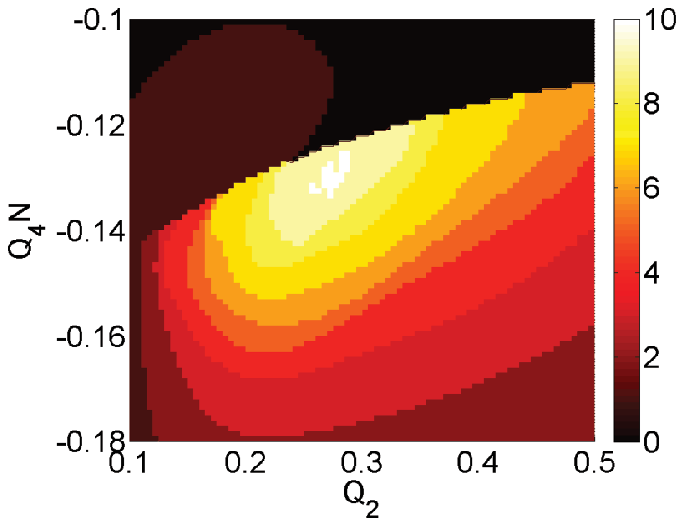}
\caption{Upper and middle figures: Time evolution of the expectation values 
of the local relative particle number operators, 
$<\hat{n}_i> \equiv <\hat{N}_i>/N$, for localized initial quantum states 
obtained from superpositions of eigenstates in the lowest-energy band 
of the $f=4$-site lattice (periodic boundary conditions), 
after imprinting a phase gradient $\theta=0.1$ (upper figures) and 
$\theta=1$ (middle figures). The number of particles is $N=20$. 
Parameter values in the eBH Hamiltonian (\ref{johansson:eq:eBHM}) are 
$Q_2=-Q_5 N = 0.3$, and $Q_4 N = -0.16$ (left figures) 
and  $Q_4 N = -0.12$ (right figures), respectively.
Lower figure: The number of sites a state with $\theta=1$ can travel before
the maximum local population expectation values have decayed to 0.4 at times 
when they are equal, $<\hat{n}_i> =<\hat{n}_{i+1}>$ (i.e., intersection points 
in (d)), plotted as functions of $Q_2$ and $Q_4 N$ while keeping  
$Q_5 N = -Q_2$ corresponding to the classical 2-site compacton condition. 
$f=4, N=20$. Adapted from Ref.~\cite{jason2013}. }
\label{fig:johansson-figure11}
\end{center}
\end{figure}
In Fig.~\ref{fig:johansson-figure11} we illustrate, 
for a system of $N=20$ particles, the quantum dynamics of 
Eq.~(\ref{johansson:eq:eBHM}) resulting 
from using such linear combinations of lowest-energy eigenstates as initial
states, 
after applying an initial ``kick'' in 
order to 
induce a directed mobility of these states. 
Analogously to kicking a classical 
soliton/breather, a phase gradient is imprinted by acting on the state 
with the phase operator $\exp(\I \theta \sum_j j \hat{N}_j)$, which corresponds 
to imposing a classical phase gradient  $\theta$ as discussed, e.g., 
in  Ref.~\cite{mishmash2009}. Figures (a)-(d)
illustrate a typical scenario in a regime where the classical ground 
state is a two-site compacton. Away from the immediate neighbourhood 
of the stability exchange regime ((a) and (c)), we can clearly identify signs 
of the classical PN barrier in the quantum dynamics: in (a), when the kick 
is too small to overcome the PN barrier, the site population expectation 
values exhibit small oscillations
around their initial equal distribution, slowly decaying due 
to the quantum tunneling, while in (c), when the kick of the 
same initial state is strong enough for overcoming the barrier, the main 
population starts to move around the lattice. On the other hand, for 
parameter values close to the classical 
stability exchange region ((b) and (d)) where the classical compacton becomes 
mobile already for very small kicks, 
the quantum tunneling times decrease as discussed 
above, and therefore, for small kicks as in (b), 
the quantum spreading takes over before the soliton has had the time 
to translate even one site. However, for larger kicks (d), the time scale 
of the classical movement becomes short enough to separate from the quantum 
time scale, and therefore the soliton population may move in a classical-like 
way for rather long distances. 

Thus, the existence of a classical stability exchange regime can be said 
to play a ``double game'' for the quantum mobility of localized initial states. 
On one hand, it lowers 
the PN barrier making the movement of highly localized states at all possible. 
On the other hand, it decreases drastically the quantum tunneling times, so 
that only solitons with sufficiently high velocities to separate from 
the quantum time-scale can move coherently for longer distances. Results 
showing the dependence of the ``fast'' mobility when varying the model 
parameters are summarized in the lower plot in 
Fig.~\ref{fig:johansson-figure11}. The whitest part corresponds to the 
optimal mobility regime, where the initially 2-site compacton-like soliton 
may travel for 10 sites before 
its maximum population expectation values at inter-site positions have 
decreased to 80\% of their initial values. The sharp transition to a dark 
region when increasing $Q_4$ is a direct counterpart of the classical 
stability exchange: for larger $Q_4$, the ground-state is on-site rather than 
inter-site centered, and thus the initial state in this regime 
bears no resemblance to a 2-site compacton. 

There are also alternative ways to construct localized quantum states which 
correspond 
to certain well-defined stationary states in the classical limit, such 
as the use of SU($f$) coherent states 
(see, e.g., Ref.~\cite{buonsante2005} for definition and discussion). 
As discussed in Ref.~\cite{jason2013}, we may describe a 2-site compacton
as an SU(2) coherent state, then kick it by applying 
the phase operator, and use it as initial conditions for the quantum 
simulations analogously to above. The results are similar (the reader 
is referred to Ref.~\cite{jason2013} for details), which shows that the 
conclusions above are not critically dependent on the specific choice of 
a quantum ``compacton-like'' initial state. One advantage with using 
the SU($f$) construction is, that it works equally well in the regimes 
where the 2-site compacton is not the ground state. Thus, we could kick
also an unstable 2-site SU($f$) 
compacton and observe good large-velocity mobility
close to the stability-exchange regime (essentially, we obtain a picture 
similar to the lower plot in Fig.~\ref{fig:johansson-figure11} but without
the sharp transition to the dark area in the upper part). 

\section{Conclusion}

We hope the reader has enjoyed this brief review about the role of the 
concepts of PN potential and barrier for describing breather mobility, 
focusing mainly on the progress from the last 10-15 years on mobility 
of strongly localized modes, mobility in two-dimensional lattices, 
moving breathers in dissipative lattices with intrinsic gain, and mobility 
of strongly localized quantum breathers. We certainly did not make any 
attempt to give a complete review on the topic of moving breathers 
(that would in itself require a whole volume!), and we are aware of many 
important references that have been omitted. Instead, our main aim was to 
collect a number of different results which have previously appeared 
scattered in the literature into a common framework; although they address 
seemingly quite different physical systems such as the classical DNLS model 
with various modifications in 1D and 2D, flat-band modes, discrete 
Ginzburg-Landau models, and the quantum extended Bose-Hubbard model, they 
all share a central core of analyzing mobility of strongly localized 
modes in terms of the PN potential concept. Obviously, the description 
in terms of PN potentials is certainly not the only method needed in order to 
get a complete understanding of the very complex problem of moving breathers 
(in particular, the more mathematically oriented reader can be directed 
to Chapter 5 of Ref.~\cite{pelinovsky2011} for a nice survey of various 
approaches used in more rigorous contexts). However, we might dare to say 
that without using these concepts, not much physical insight into the 
mechanisms by which localized excitations can be translated in any lattice 
(or, more generally, periodic potential) would have been reached. We are 
also certain that many more future applications will appear!


\section*{Acknowledgments} 
The main part of the 
work presented in Secs.~\ref{johansson:sec:1D}-\ref{johansson:sec:gain}
was performed by one of us (M.J.) 
in collaboration with, in chronological order, 
Michael {\"Oster} (Sec.~\ref{johansson:sec:1D}), Rodrigo Vicencio 
(Secs.~\ref{johansson:subsec:2Dsat}-\ref{johansson:subsec:kagome}), 
Uta Naether (Sec.~\ref{johansson:subsec:2Dsat}), and Jaroslaw Prilepsky and 
Stanislav Derevyanko (Sec.~\ref{johansson:sec:gain}).
M.J. is particularly grateful to the collaborators for producing the original 
versions of all figures, as well as the underlying numerical simulations, 
in Secs.~\ref{johansson:sec:1D}-\ref{johansson:sec:2D}. M.J.\ is also 
very grateful to Rodrigo Vicencio and Stanislav Derevyanko for their kind, 
repeated invitations to visit the Nonlinear Optics Group, Universidad de 
Chile, and the School of Engineering and Applied Science, Aston University, 
respectively, during which a large part of this work was realized. 
Much of the work presented here stems from original ideas of Serge Aubry 
and Sergej Flach, who are especially thanked by M.J.\ for many enlightening 
discussions on breather mobility. Among the many other colleagues who 
contributed with suggestions on various occasions, M.J.\ would 
like to give a special 
mentioning to Thierry Cretegny, from whom he first learned many of the 
intricate scenarios for breather mobility, and to Yaroslav Zolotaryuk 
for first pointing out to him the similarities with nontrivial PN potentials 
for kinks. Finally, Chris Eilbeck pioneered both topics of breather mobility 
and quantum breathers, and we are very grateful for many discussions
during the years. Parts of this work were supported by the Swedish Research 
Council.
\\ \mbox{}\\


\newcommand{\noopsort}[1]{} \newcommand{\printfirst}[2]{#1}
  \newcommand{\singleletter}[1]{#1} \newcommand{\switchargs}[2]{#2#1}

\end{document}